\newcommand*{\ATLASLATEXPATH}{}
\documentclass[cernpreprint, texlive=2016, UKenglish]{\ATLASLATEXPATH atlasdoc}
 
\usepackage{\ATLASLATEXPATH atlaspackage}
\usepackage{\ATLASLATEXPATH atlasbiblatex}
 
\usepackage{\ATLASLATEXPATH atlasphysics}
 
\graphicspath{{logos/}{figures/}}
 
\usepackage{ANA-STDM-2017-24-PAPER-defs}
\AtlasTitle{Measurement of fiducial and differential $W^+W^-$ production cross-sections at $\sqrt{s}=$13~TeV with the ATLAS detector}

\AtlasAbstract{
A measurement of fiducial and differential cross-sections for $W^+W^-$ production in proton--proton collisions at $\sqrt{s}=$13~TeV with the ATLAS experiment at the Large Hadron Collider using data corresponding to an integrated luminosity of $36.1$~fb$^{-1}$ is presented. Events with one electron and one muon are selected, corresponding to the decay of the diboson system as $WW\rightarrow e^{\pm}\nu\mu^{\mp}\nu$. To suppress top-quark background, events containing jets with a transverse momentum exceeding 35~GeV are not included in the measurement phase space.
The fiducial cross-section, six differential distributions and the cross-section as a function of the jet-veto transverse momentum threshold are measured and compared with several theoretical predictions.
Constraints on anomalous electroweak gauge boson self-interactions are also presented in the framework of a dimension-six effective field theory.
}
 
\author{The ATLAS Collaboration}
 
\AtlasRefCode{STDM-2017-24}

\PreprintIdNumber{CERN-EP-2019-055}
 
\AtlasJournalRef{Eur. Phys. J. C 79 (2019) 884}
 
\AtlasDOI{10.1140/epjc/s10052-019-7371-6}

\hypersetup{pdftitle={ATLAS document},pdfauthor={The ATLAS Collaboration}}
 
\begin{document}
 
\maketitle
 
\tableofcontents

\section{Introduction}
\label{sec:intro}
 
The measurement of the production of \Wboson-boson pairs through interactions of quarks and gluons
probes the electroweak (EW) gauge structure of the Standard Model (SM) and allows further tests of the strong interaction between quarks and gluons.
The \WW production process is also important as it constitutes large irreducible backgrounds in searches for physics beyond the SM and to $H\rightarrow WW^*$ production.
Its large production cross-section combined with the large sample of proton--proton ($pp$) collision data delivered by the Large Hadron Collider (LHC), enables this process to be studied differentially with a better statistical precision than was possible in previous measurements.
 
The first measurements of \WW production were carried out at the LEP electron--positron collider~\cite{schael:2013ita}. At the Tevatron this process was measured in proton--antiproton collisions by the CDF~\cite{Abe:1996dw,Aaltonen:2009aa} and D\O{}~\cite{Abazov:2009ys} Collaborations. In $pp$ collisions at the LHC, \WW production cross-sections were determined for centre-of-mass energies of $\sqrt{s}=7$~\TeV\ and $\sqrt{s}=8$~\TeV\ by the ATLAS~\cite{ww7tev,STDM-2013-07} and CMS~\cite{CMS-SMP-12-005,Khachatryan:2015sga} Collaborations. In addition, a dedicated measurement of the $\WW+1$-jet final state was carried out by the ATLAS Collaboration~\cite{STDM-2015-24} at $\sqrt{s}=8$~\TeV.
At $\sqrt{s}=13$~\TeV, the total cross-section for \WW production was measured by the ATLAS Collaboration~\cite{STDM-2015-20}, albeit only for the small 2015 data sample, which did not allow any differential studies.
 
The cross-section measurements at $\sqrt{s}=7$ and $\sqrt{s}=8$~\TeV\ revealed discrepancies between data and theory that have since been addressed through the inclusion of higher-order corrections in perturbative quantum chromodynamics (QCD)~\cite{Gehrmann:2014fva,Meade:2014fca,Jaiswal:2014yba,Monni:2014zra,Grazzini:2016ctr,Dawson:2016ysj}. This has remedied the mismatch between the total measured and predicted cross-sections, but some discrepancies in the differential distributions persist.
The high-energy behaviour of the \WW cross-section and the angular distributions of the \WW decay products could be affected by new physical phenomena at higher partonic centre-of-mass energies, such as EW doublet or triplet scalars~\cite{deBlas:2017xtg,Henning:2014wua} or degenerate and non-degenerate top-quark superpartners (stops) in supersymmetry (SUSY) scenarios~\cite{Ellis:2018gqa, Drozd:2015kva}. These specific models can be constrained by their contribution to dimension-six operators in an effective Lagrangian at tree level~\cite{deBlas:2017xtg}. At lower partonic centre-of-mass energies, \WW production can also be used to provide complementary constraints on compressed EW SUSY scenarios with low stop masses~\cite{tattersal}.
 
\begin{figure}[htp]
\centering
\includegraphics[width=0.24\textwidth]{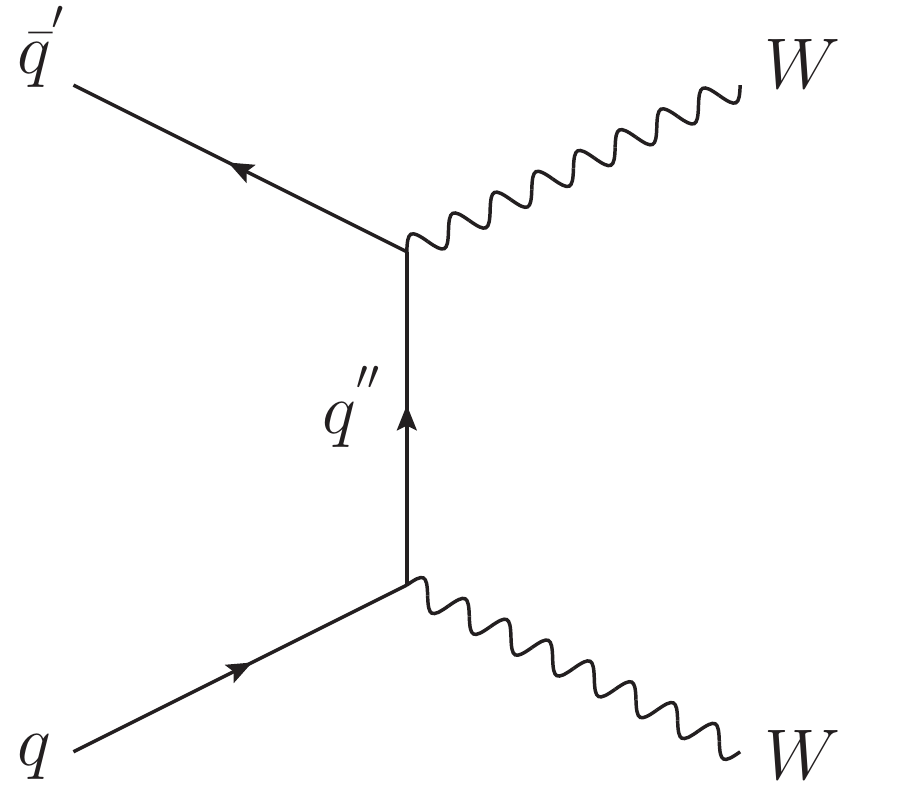}
\includegraphics[width=0.24\textwidth]{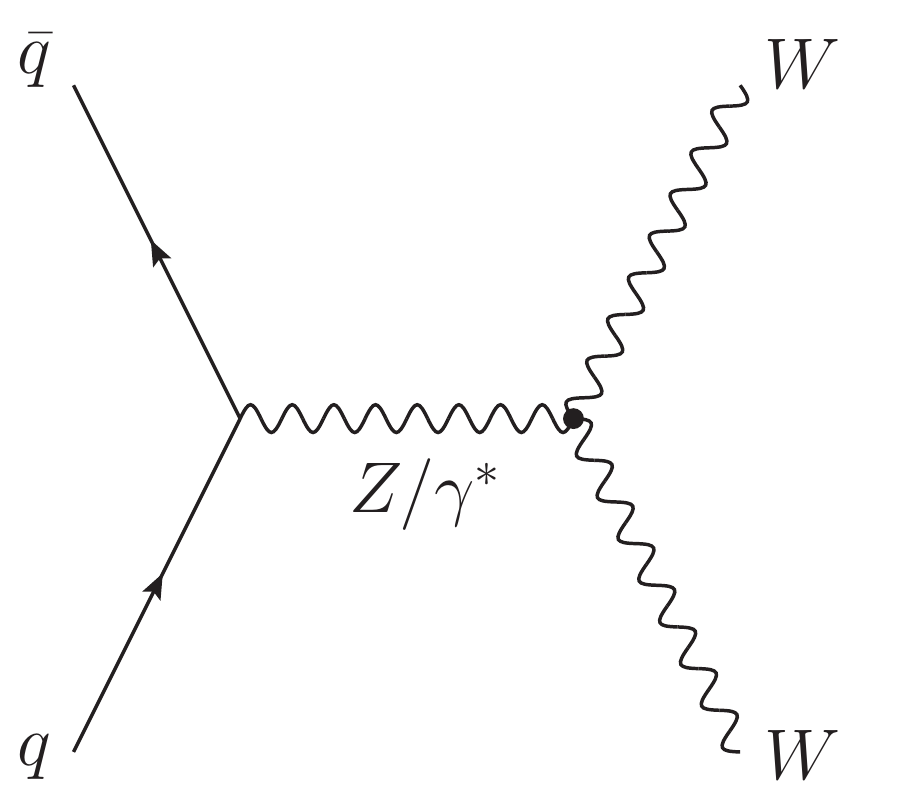}
\includegraphics[width=0.24\textwidth]{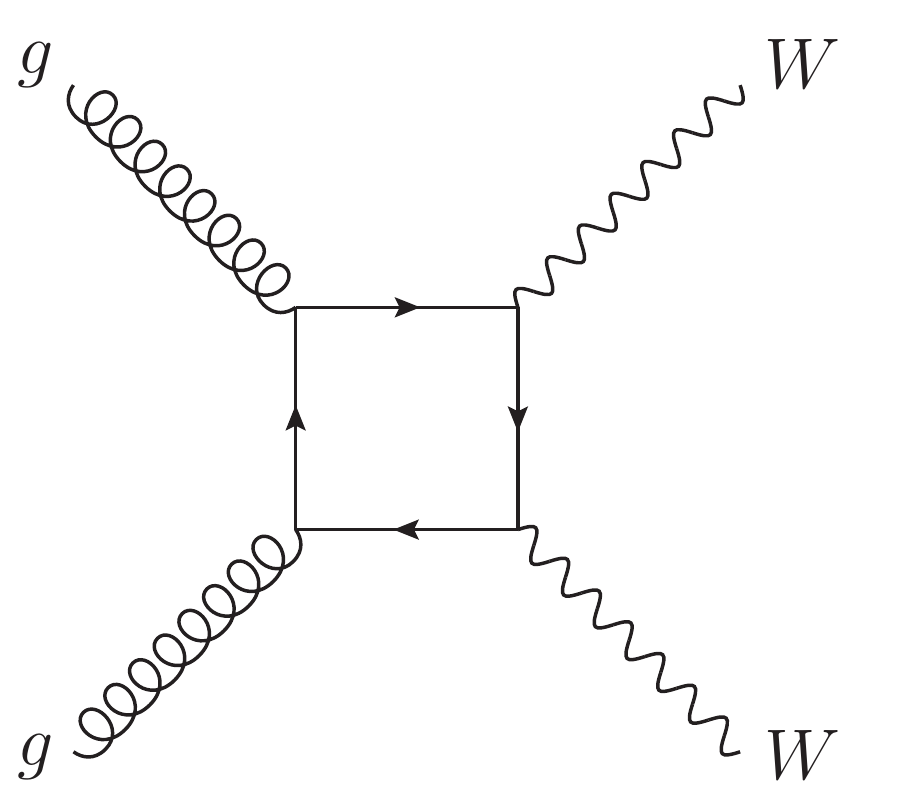}
\includegraphics[width=0.24\textwidth]{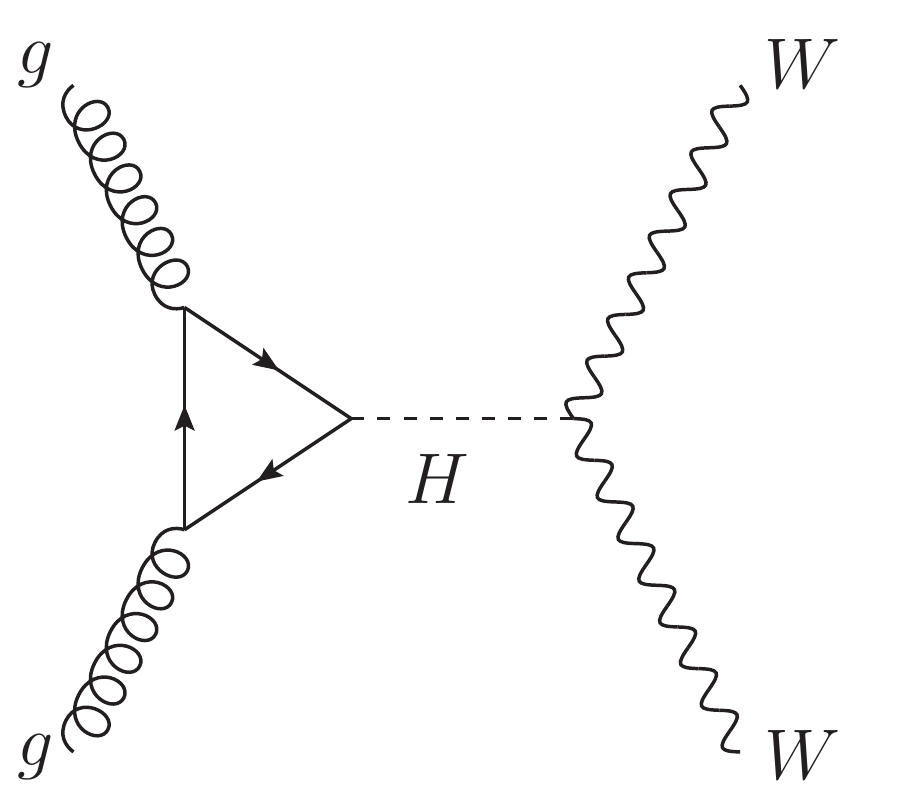}
\caption{Feynman diagrams for SM $WW$ production at tree level (from left to right): \qqbar initial-state $t$-channel, \qqbar initial-state $s$-channel, $gg$ initial-state non-resonant and $gg$ initial-state resonant production. The $s$-channel production contains the $WWZ$ and $WW\gamma$ triple-gauge-coupling vertices. The gluon--gluon fusion processes are mediated either by a quark loop (\ggtoWW) or the resonant production of a Higgs boson with subsequent decay into $WW$ (\ggtoHtoWW). }
\label{fig:Intro_WWprodFeynmanDiagrams}
\end{figure}

The \WW signal is composed of two leading sub-processes: \qqtoWW production\footnote{The notation \qqtoWW is used to include both the \qqbar and $qg$ initial states for $\Wboson\Wboson$ production.} (in the $t$- and $s$-channels) and gluon--gluon fusion production (both non-resonant \ggtoWW and resonant \ggtoHtoWW). \figref{Intro_WWprodFeynmanDiagrams} shows representative sub-processes.
To allow for a proper treatment and inclusion of the interference, which is especially relevant in the
tails of kinematic distributions, the resonant production is kept as part of the signal. The fiducial phase space is defined to be orthogonal to the $H\rightarrow WW$ measurements by the ATLAS Collaboration~\cite{HIGG-2013-13, Aaboud:2018jqu} using a requirement on the dilepton invariant mass.
Therefore the Higgs boson contribution included in the signal definition is dominated by off-shell production and interference effects. The production of two $W$ bosons from the decay of top--antitop quark pairs is not considered part of the signal.
 
The different sub-processes for \WW production are known theoretically at different orders in the strong coupling constant $\alpha_{\mathrm{s}}$. The \qqtoWW production cross-section is known to $\mathcal{O}(\alpha_{\rm s}^2)$, next-to-next-to-leading order (NNLO)~\cite{Gehrmann:2014fva,Grazzini:2016ctr}.
Recently, also a NNLO prediction matched to a parton shower has become available~\cite{WWNNLOPS,Grazzini:2016ctr,WWMiNLO}.
The non-resonant \ggtoWW production cross-section is known to $\mathcal{O}(\alpha_{\rm s}^3)$, next-to-leading order (NLO)~\cite{Caola:2015rqy}, and its interference with the resonant \ggtoWW production cross-section is known to $\mathcal{O}(\alpha_{\rm s}^2)$.
 
This paper presents a measurement of the fiducial cross-section for \WW\ production at $\sqrt{s}=13~\tev$ using $pp$ collision data recorded in 2015 and 2016 by the ATLAS experiment, corresponding to an integrated luminosity of \intlumi.
The $WW\rightarrow e^\pm\nu\mu^\mp\nu$ decay channel is studied (denoted in the following by \WWemu).
The measurement is performed in a phase space close to the geometric and kinematic acceptance of the experimental analysis. This includes a veto on the presence of jets with transverse momenta (\pt) above a series of thresholds, with a $\pt=35~\gev$ threshold used as a baseline.
Measuring the fiducial cross-section as a function of the jet veto \pt threshold provides an indirect measure of the jet \pt spectrum in $WW$ events, without removing the jet veto that is necessary for background suppression.
 
Six differential distributions involving kinematic variables of the final-state charged leptons are measured in the baseline phase space. Three of them characterize the energy of the process: the transverse momentum of the leading lepton \ptlzero, the invariant mass of the dilepton system \mll\ and the transverse momentum of the dilepton system \ptll. Three further distributions probe angular correlations and the spin state of the \WW\ system. These are the rapidity of the dilepton system \yll, the difference in azimuthal angle between the decay leptons \dphill, and \costheta defined as:
\begin{linenomath*}
\begin{equation*}
\abscostheta\ = \left|\tanh\left(\frac{\Delta \eta_{e\mu}}{2} \right) \right|~,
\end{equation*}
\end{linenomath*}
where $\Delta \eta_{e\mu}$ is the difference between the pseudorapidities of the leptons.\footnote{ATLAS uses a right-handed coordinate system with its origin at the nominal interaction point in the centre of the detector and the $z$-axis coinciding with the axis of the beam pipe. The $x$-axis points from the interaction point to the centre of the LHC ring, and the $y$-axis points upward.  The pseudorapidity is defined in terms of the polar angle $\theta$ as $\eta=-\ln \tan(\theta/2)$, and $\phi$ is the azimuthal angle around the beam pipe relative to the $x$-axis.
The angular distance is defined as $\Delta R=\sqrt{(\Delta \eta)^2+(\Delta \phi)^2}$.
Transverse energy is computed as $E_{\textrm{T}} = E \cdot \sin\theta$.}
This variable is longitudinally boost-invariant and sensitive to the spin structure of the produced diparticle pairs as discussed in Ref.~\cite{Barr:2005dz}.
The unfolded \ptlzero\ distribution  is used to set limits on anomalous triple-gauge-boson couplings,
since this distribution was identified as the most sensitive to the effect of these couplings.
 
 
\section{ATLAS detector}
\label{sec:detector}
 
The ATLAS detector~\cite{PERF-2007-01} at the LHC is a multipurpose particle detector with a forward--backward symmetric cylindrical geometry and nearly 4$\pi$ coverage in solid angle.
It consists of inner tracking devices surrounded by a superconducting solenoid, electromagnetic (EM) and hadronic calorimeters, and a muon spectrometer.
The inner detector (ID) provides charged-particle tracking in the pseudorapidity region $|\eta|<2.5$ and vertex reconstruction.
It comprises a silicon pixel detector, a silicon microstrip tracker, and a straw-tube transition radiation tracker.
The ID is placed inside a solenoid that produces a 2\,T axial magnetic field.
Lead/liquid-argon (LAr) sampling calorimeters provide EM energy measurements with high granularity.
A steel/scintillator-tile hadronic calorimeter covers the central pseudorapidity range $|\eta|<1.7$.
The endcap and forward regions are instrumented with LAr calorimeters for both the EM and hadronic energy measurements up to $|\eta|=4.9$.
The muon spectrometer (MS) is operated in a magnetic field provided by air-core superconducting toroids and includes tracking chambers
for precise muon momentum measurements up to $|\eta|=2.7$ and trigger chambers covering the range $|\eta|<2.4$.
 
A two-level trigger system~\cite{TRIG-2016-01} selects the events used in the analysis.
The first level is implemented in custom electronics, while the second trigger level is a flexible software-based system.
 
\section{Data and simulated event samples}
\label{sec:dataMCsamples}
 
The data were collected at a centre-of-mass energy of 13~\TeV\ during 2015 and 2016, and correspond to an integrated luminosity of \intlumi. Only high-quality data with all detectors in normal operating conditions are analysed.
The average number of interactions per bunch crossing was estimated to be $\left<\mu\right>=24$.
 
Simulated event samples are used for most of the background estimates, for the correction of the signal yield due to detector effects, and for comparison with the measured cross-sections.
 
The \WW signal was modelled using the NLO perturbative QCD \POWHEGBOXV{v2} event generator~\cite{Nason:2004rx, Frixione:2007vw, Alioli:2010xd, Melia:2011tj, Nason:2013ydw} for $q\bar{q}$ initial states.
The $gg\rightarrow WW$ contribution was generated using the \sherpa~2.1.1+\openloops\ framework~\cite{Hoeche:2012yf, jetvetoNNLO} at leading order (LO) with up to one additional parton and includes non-resonant and resonant Higgs boson production and interference terms. The \sherpa~2.1.1+\openloops\ framework also allows these contributions to be generated and studied separately.
In both cases, the \ctten~\cite{CTEQ10} parton distribution functions (PDF) were used.
\POWHEGBOX was interfaced to \PYTHIAV{8.210}~\cite{Sjostrand:2014zea} for the modelling of parton showers and hadronization as well as underlying-event simulation, using the \aznlo~\cite{AZNLO:2014} set of tuned parameters (`tune') and the \ctsix ~\cite{Pumplin:2002vw} PDF set.
\SHERPA used its own parton shower, fragmentation and underlying-event model.
Alternative signal samples for the quark-induced production were generated using \POWHEGBOX interfaced to \herwigpp\ 2.7.1~\cite{Bahr:2008pv} with the UEEE5 tune~\cite{Gieseke:2012ft}, and using the \sherpa~2.2.2 generator with its own model for parton showering, hadronization and the underlying event.
The \SHERPAV{2.2.2} prediction was obtained at NLO with up to one additional parton emission and up to three at LO and employs the NNPDF3.0~\cite{Ball:2014uwa}  PDF set.
The \WW signal predictions were normalized to the NNLO cross-section~\cite{Gehrmann:2014fva}; the $gg\rightarrow WW$ process was normalized to its inclusive NLO cross-section~\cite{Caola:2015rqy}.
 
The background processes considered are: top-quark pair production (\ttbar), associated production of a top quark with a $W$ boson ($Wt$), single vector-boson production ($W$ or $Z$, in association with jets), multijet production, other diboson production ($WZ$, $ZZ$, $W\gamma$ and $Z\gamma$) and triboson production ($WWW$, $WWZ$, $WZZ$ and $ZZZ$), where $Z$ stands for $Z/\gamma^{*}$.
 
For the generation of \ttbar and $Wt$ processes at NLO, \POWHEGBOXV{v2}~\cite{Frixione:2007nw} and \POWHEGBOXV{v1}~\cite{Nason:2004rx} respectively were used with the {CT10} PDF set.
For the parton shower, hadronization and underlying event, simulated events were interfaced to \pythia\ {8.186} for \ttbar and \pythia\ {6.425}~\cite{Sjostrand:2000wi} for single-top production, using the \afourteen\ tune~\cite{ATL-PHYS-PUB-2014-021} and the {Perugia 2012}~\cite{Skands:2010ak} tune, respectively.
The top-quark mass was set to 172.5~\GeV.
In the \ttbar sample, the $h_{\mathrm{damp}}$ parameter that regulates the high-\pT emission, against which the \ttbar system recoils, was set to 1.5 times the top-quark mass following studies reported in Ref.~\cite{topMCs}.
Alternative samples were generated with different settings to assess the uncertainty in modelling top-quark events.
To estimate uncertainties in additional QCD radiation in top-quark processes, a pair of samples was produced with the alternative sets of \afourteen\ (\ttbar) or {Perugia 2012} ($Wt$) parameters for higher and lower radiation, as well as with different renormalization and factorization scales which were both varied either by a factor of 2 or 0.5.
For the higher-radiation samples, the value of the $h_{\mathrm{damp}}$ parameter was doubled.
Two alternative Monte Carlo (MC) programs were used to estimate the impact of the choice of hard-scatter generator and hadronization algorithm in top-quark events; for each of these samples one of the two components was replaced by an alternative choice. The alternative choices are \MGMCatNLOV{2.3}~\cite{Alwall:2014hca} for the hard-scatter generator and \HERWIGV{7}~\cite{Bellm:2015jjp} (\herwigpp\ 2.7.1) for the hadronization algorithm in \ttbar ($Wt$) events.
In addition, the modelling of the overlap at NLO between $Wt$ and $t\bar{t}$ diagrams~\cite{Frixione:2008yi} was studied.
The effect was assessed by generating $Wt$ events with different schemes for overlap removal using the \POWHEGBOX event generator interfaced to \PYTHIAV{6.425} for the simulation of parton showering and non-perturbative effects.
The top-quark events were normalized using the NNLO+next-to-next-to-leading-logarithm (NNLL) QCD cross-section~\cite{ttbarXsec7} for \ttbar, and the NLO+NNLL cross-section~\cite{Kidonakis:2010ux} for $Wt$ production.
 
The $Z$+jets process (with $Z\rightarrow ee/\mu\mu/\tau\tau$) was modelled using \SHERPAV{2.2.1}~\cite{Sherpa} with the  NNPDF3.0 PDF set.
This process was calculated with up to two additional partons at NLO and up to four additional partons at LO.
The $W$+jets and alternative $Z$+jets events were produced with the \POWHEGBOX generator at NLO accuracy using the {CT10} PDF set, interfaced to \PYTHIAV{8.186} for parton showering, hadronization and the underlying event. As in the \WW samples, the \aznlo\ tune was used for the underlying event together with the \ctsix\ PDF set.
The $Z$+jets and $W$+jets events were normalized using their respective NNLO cross-section calculations~\cite{Anastasiou:2003ds}.
 
The background from diboson production processes ($WZ$, $ZZ$, $W\gamma$ and $Z\gamma$) was simulated using the \SHERPAV{2.2.2} generator with the NNPDF 3.0 PDF set. The samples include up to one additional parton emission at NLO and up to three at LO.
Alternative samples for $WZ$ and $ZZ$ processes were produced using the same \POWHEGBOX{}+\PYTHIAV{8} set-up as the $q\bar{q}$-initiated \WW signal samples discussed above.
The background from triboson production was modelled using the \sherpa+\openloops\ generator with the \ctten\ PDF set, calculated at NLO for inclusive production and including up to two hard parton emissions at LO.
The $WZ$, $ZZ$ and triboson samples produced with \sherpa\ were normalized to the cross-section calculated by \sherpa, with hard parton emissions at NLO or LO as discussed, and thus already capturing some of the NNLO effects.
The $WZ$ and $ZZ$ backgrounds simulated with \POWHEGBOX were normalized to their NNLO cross-sections~\cite{Grazzini:2016swo,Grazzini:2017ckn,Grazzini:2015hta,Cascioli:2014yka,Kallweit:2018nyv}.
 
\evtgen\ {1.2.0}~\cite{evtgen} was used for the properties of the bottom and charm hadron decays after hadronization in all samples generated with \POWHEGBOX and \MGMCatNLO.

Additional interactions in the same or nearby bunch crossings (pile-up) were simulated using \PYTHIAV{8.186} using the A2 tune~\cite{ATL-PHYS-PUB-2012-003} and the MSTW2008LO PDF~\cite{Martin:2009iq} set and were overlaid on the simulated signal and background events.
 
All simulated event samples were produced using the ATLAS simulation infrastructure~\cite{SOFT-2010-01}, using the full \GEANT4~\cite{Geant4} simulation of the ATLAS detector.
Simulated events were then reconstructed with the same software as used for the data and were corrected with data-driven correction factors to account for differences in lepton and jet reconstruction and identification between data and simulation.
These corrections are of the order of 1--3\%.

 
\section{Event reconstruction and selection}
\label{sec:selection}

The \WW event candidates are selected by requiring each event to contain exactly one electron and exactly one muon of opposite charge, each passing the selections described below. Events with a same-flavour lepton pair are not used because they have a larger background from the Drell--Yan process.
 
Candidate events are required to have at least one vertex with at least two associated tracks with $\pt > 400$~\mev. The vertex with the highest $\sum\pt^2$ of the associated tracks is considered to be the primary vertex.
 
\subsection{Trigger}
Candidate events were recorded by either a single-muon or a single-electron trigger that imposed a minimum lepton transverse momentum threshold that varied during data-taking.
The \pt\ threshold of the leptons required by triggers in 2015 was 24 \GeV\ for electrons and 20 \GeV\ for muons, both satisfying loose isolation requirements.
Due to the higher instantaneous luminosity in 2016 the trigger threshold was increased to 26 \GeV\ for both the electrons and the muons,
and more restrictive isolation for both the leptons as well as more restrictive identification requirements for electrons were applied. Additionally, single-lepton triggers with higher \pt\ thresholds but with no isolation or with loosened identification criteria were used to increase the efficiency.
The trigger efficiency for events satisfying the full selection criteria described below is about 99\% and is determined using a simulated signal sample that is corrected to reflect the data efficiencies with corrections measured using $Z\rightarrow ee$~\cite{Aaboud:2019ynx} and $Z\rightarrow \mu\mu$~\cite{PERF-2015-10} events. These data-driven corrections are of the order of 2\% with permille level uncertainties.

\subsection{Leptons}
 
Electron candidates are reconstructed from the combination of a cluster of energy deposits in the EM calorimeter and a track in the ID~\cite{Aaboud:2019ynx}. Candidate electrons must satisfy the {\it TightLH} quality definition described in Ref.~\cite{Aaboud:2019ynx}.
Signal electrons are required to have $\eT > 27$~\GeV\ and the pseudorapidity of electrons is required to be  $|\eta| < 2.47$, excluding the transition region between the barrel and endcaps in the LAr calorimeter ($1.37 < |\eta| < 1.52$).
In addition, a requirement is added to reject electrons that potentially stem from photon conversions to reduce the $W\gamma$ background~\cite{Aaboud:2019ynx}. This uses a simple classification based on the candidate electron’s $E/p$ and $\pT$,
the presence of a hit in the pixel detector, and the secondary-vertex information, to determine whether the electron could also be considered as a photon candidate and rejected.
 
Muon candidates are reconstructed by combining a track in the ID with a track in the MS~\cite{PERF-2015-10}. The {\it Medium} quality criterion, as defined in Ref.~\cite{PERF-2015-10}, is applied to the combined tracks.
Signal muons are required to have $\pT>27$~\GeV~and $|\eta| <$ 2.5.
 
Leptons are required to originate from the primary vertex.
The longitudinal impact parameter of each lepton track, calculated relative to the primary vertex and multiplied by $\sin\theta$ of the track, is required to be smaller than 0.5 mm.
Furthermore, the significance of the transverse impact parameter, defined by the transverse impact parameter ($d_0$) of a lepton track relative to the beam line, divided by its estimated uncertainty ($\sigma_{d_0}$), is required to satisfy $|d_0/\sigma_{d_0}| <3.0~(5.0)$ for muons (electrons).
Leptons are also required to be isolated using information from ID tracks and energy clusters in the calorimeters in a cone around the lepton. The expected isolation efficiency is at least 90\% (99\%) at a \pt of 25 (60)~\GeV~using the {\it Gradient} working point defined in Refs.~\cite{PERF-2015-10,Aaboud:2019ynx}.
 
\subsection{Jets}
Jet candidates are reconstructed within the calorimeter acceptance using the anti-\kt~jet clustering algorithm~\cite{antikt-jet} using the FastJet code~\cite{Cacciari:2011ma} with a radius parameter of $R$ = 0.4, which combines clusters of topologically connected calorimeter cells~\cite{ATL-PHYS-PUB-2015-036, PERF-2014-07}. The jet energy is calibrated by applying a \pt- and $\eta$-dependent correction derived from MC simulation with additional corrections based on data~\cite{PERF-2016-04}. As part of the jet energy calibration a pile-up correction based on the concept of jet area is applied to the jet candidates~\cite{ATLAS-CONF-2017-065}.
Jets are required to have a pseudorapidity $|\eta|<4.5$.
 
The jet-vertex-tagger (JVT) technique~\cite{PERF-2014-03} is used to separate hard-scatter jets from pile-up jets within the acceptance of the tracking detector by requiring a significant fraction of the jets' summed track \pt to come from tracks associated with the primary vertex.
For jets with $2.5<|\eta|<4.5$, a forward-JVT selection is applied to suppress pile-up jets~\cite{PERF-2016-06}.
 
Candidate jets are discarded if they are within a cone of size $\Delta R = 0.2$ around an electron candidate, or if they have fewer than three associated tracks and are within a cone of size $\Delta R = 0.2$ around a muon candidate.
However, if a jet with three or more associated tracks is within a cone of size $\Delta R = 0.4$ around a muon candidate, or any jet is within a region $0.2< \Delta R < 0.4$ around an electron candidate, the corresponding electron or muon candidate is discarded.
 
Within the ID acceptance, jets originating from the fragmentation of $b$-hadrons ($b$-jets) are identified using a multivariate algorithm (MV2c10 BDT)~\cite{ATL-PHYS-PUB-2015-022, ATL-PHYS-PUB-2015-039}. The chosen operating point has an efficiency of 85\% for selecting jets containing $b$-hadrons, as estimated from a sample of simulated \ttbar events and validated with data~\cite{ATL-PHYS-PUB-2015-039}.

\subsection{Missing transverse momentum}
 
The missing transverse momentum is computed as the negative of the vectorial sum of the transverse momenta of tracks associated with jets and muons, as well as tracks in the ID that are not associated with any other component. The \pT of the electron track is replaced by the calibrated transverse momentum of the reconstructed electron~\cite{PERF-2014-04}. This definition has been updated for Run 2 data-taking conditions~\cite{PERF-2016-07}, and denoted by \ptmissvec with its absolute value denoted by \ptmiss.
The tracks are required to be associated with the primary vertex and to satisfy the selection criteria described in Ref.~\cite{PERF-2016-07}.
 
The \ptmiss takes advantage of the excellent vertex resolution of the ATLAS detector and gives a missing transverse momentum estimate that is robust in the presence of pile-up, but it neglects the contribution of neutral particles, which do not form tracks in the ID. The pseudorapidity coverage of \ptmiss is also limited to the tracking volume of $|\eta| < 2.5$, which is smaller than the calorimeter coverage of $|\eta| < 4.9$.
For events without any reconstructed jets, the \ptmiss provides a small improvement of the $E_{\textrm{T}}^{\textrm{miss}}$ resolution compared with the standard reconstruction algorithms~\cite{PERF-2016-07}.

\subsection{Signal region definition}
The signal region (SR), in which the measurement is performed, is defined as follows.
To reduce the background from other diboson processes, events are required to have no additional electrons or muons with \pt $>$ 10~\GeV\ fulfilling loosened selection criteria.
For this looser selection, the {\it GradientLoose} isolation requirement~\cite{PERF-2015-10,Aaboud:2019ynx} is used for both the electrons and the muons, which has an expected isolation efficiency of at least 95\% (99\%) at a \pT of 25 (60)~\GeV. Moreover, a less stringent {\it MediumLH} requirement~\cite{Aaboud:2019ynx} is applied for electron identification.
 
To suppress the background contribution from top-quark production, events are required to have no jets with \pt$>$ 35~\GeV\ and $|\eta|<$ 4.5, and no $b$-jets with \pt $>$ 20~\GeV\ and $|\eta|<$ 2.5.
The jet \pt requirement is optimized to minimize the total systematic uncertainty in the measurement. 
The additional $b$-jet veto requirement allows the background from top-quark production to be suppressed by a factor of three, while keeping 97\% of the \WW signal events.
For the remaining top-quark background events that pass all selection criteria, the $b$-jets are mainly produced outside the acceptance of the detector (\pt $<$ 20~\GeV\ or $|\eta|>$ 2.5), according to MC simulation.
 
In addition, the requirements of \ptmiss $>$ 20~\GeV\ and $\pT^{e\mu}>30$~\GeV\ suppress the Drell--Yan background contributions. A further requirement on the invariant mass of the lepton pair ($m_{e\mu} >$ 55~\GeV) reduces the $H\rightarrow WW^*$ contribution to a level below 1\% of the expected signal. This last requirement is inverted compared with the one used in the recent measurement of $H\rightarrow WW^*$ production at 13~\TeV\ by ATLAS~\cite{Aaboud:2018jqu}, making the two measurements statistically independent.
Otherwise, both measurements use similar selections for events in the 0-jet category, although with lower lepton \pt requirements in the $H\rightarrow WW^*$  analysis.
 
The lepton, jet, and event selection criteria are summarized in Table~\ref{tab:recolevelcuts_paper}.
 
\begin{table}[t!]\renewcommand{\arraystretch}{1.2}
\caption{Summary of lepton, jet, and event selection criteria for $WW$ candidate events. In the table $\ell$ stands for $e$ or $\mu$. The definitions of lepton identification and isolation are detailed in Refs.~\cite{Aaboud:2019ynx} and \cite{PERF-2015-10}.
}
\begin{center}
\begin{tabular}{l c}
\toprule
Selection requirement                            &   Selection value \\
\midrule
$ p_{\rm T}^\ell$     	&   {$>27$~\GeV} 		\\
\hline
$ \eta^{\ell}$     				&  $|\eta^{e}|<2.47$ (excluding $1.37<|\eta^{e}|<1.52$), \\
&   $|\eta^{\mu}|<2.5$  \\
\hline
Lepton identification				&  {\it TightLH} (electron), {\it Medium} (muon)                                                    \\
\hline
Lepton isolation				&  {\it Gradient} working point                \\
\hline
Number of additional leptons ($\pt>10$ \GeV)     	&   0					\\
\hline
Number of jets ($\pt>$35 \GeV, $|\eta|<$4.5) & 0					\\
\hline
Number of $b$-tagged jets ( $\pt>20$ \GeV, $|\eta|<$2.5) 	& 0 					\\
\hline
\ptmiss    					&   $>20$ \GeV 			\\
\hline
$\pT^{e\mu}$    					&   $>30$ \GeV 	\\
\hline
$m_{e\mu}$     				&   $>55$ \GeV 			\\
\bottomrule
\end{tabular}
\end{center}
\label{tab:recolevelcuts_paper}
\end{table}

 
\section{Background estimation}
\label{sec:backgrounds}
 
After applying all selection requirements described in Section~\ref{sec:selection}, the dominant background is from top-quark production. This includes $t\bar{t}$ and $W$-associated single-top production, which both yield two real leptons in the final state.
 
The non-prompt lepton background originates from leptonic decays of heavy quarks, hadrons misidentified as leptons, and electrons from photon conversions.
Such lepton-like objects are collectively referred to as fake leptons.
Events with fake leptons  are mainly due to the production of $W+$jets, $s$- and $t$-channel single-top production, both with leptonic $W$-boson decay and a jet misidentified as a lepton, or from multijet production with two jets misidentified as leptons. Other processes can contribute as well, but are negligible in the signal region. Since most of these events -- more than $98\%$ -- correspond to $W$+jets production, this background is referred to as $W$+jets background in the following.
 
Drell--Yan production of $\tau$-leptons ($Z\rightarrow\tau\tau$) can also give rise to the $e\mu$ final state.
Other diboson ($WZ$, $ZZ$, $W\gamma$ and $Z\gamma$) and triboson ($VVV$, where $V=W,\ Z$) production processes constitute a smaller background contribution. A summary table comparing the number of observed candidate events in data to the respective numbers of predicted signal and background events in the signal region can be found in Section~\ref{sec:yieldtable}.
 
\subsection{Background from top-quark production}
\label{sec:bkg_top}
 
Background from top-quark production is estimated using a partly data-driven method~\cite{TopJVSP, STDM-2013-07}, in which the top-quark contribution is extrapolated from a control region (top CR) to the signal region.
The top CR is selected by applying the $WW$ signal selection except for the $b$-jet and jet-veto requirements.
To reduce the \WW signal contamination in this control region, an additional requirement on the scalar sum of the transverse momenta of leptons and jets, $H_{\textrm{T}} > 200$~\GeV, is applied.
The remaining non-top-quark contribution estimated by MC simulation
is subtracted and the resulting number of top-quark events, $N^{\textrm{top}}_{\textrm{CR}}$,
is corrected for the $H_{\textrm{T}}$ cut efficiency, $\epsilon_{H_{\textrm{T}}}$, using top-quark MC samples.
With the efficiency for top-quark events to satisfy the jet-veto requirement, $\epsilon_{\textrm{jet-veto}}$,
the top-quark background contribution in the signal region can be calculated as:
\begin{linenomath*}
\begin{equation*}
N^{\textrm{top}}_{\textrm{SR}} = \frac{N^{\textrm{top}}_{\textrm{CR}}}{\epsilon_{H_{\textrm{T}}}} \times  \epsilon_{\textrm{jet-veto}}~.
\end{equation*}
\end{linenomath*}
The jet-veto efficiency, which mainly quantifies the fraction of top events with jets below the jet-veto and $b$-jet-veto \pt thresholds, is calculated from simulation, with an extra correction factor~\cite{TopJVSP, STDM-2013-07}:
\begin{equation}
\label{epsilon_jet_veto}
\epsilon_{\textrm{jet-veto}} = \epsilon_{\textrm{jet-veto}}^\text{MC}\times\left(\frac{\epsilon_\text{single-jet-veto}^\text{Data}}{\epsilon_\text{single-jet-veto}^\text{MC}}\right)^{\langle n_\text{jets} \rangle}
\end{equation}
where $\epsilon_\text{single-jet-veto}$ is defined as the fraction of top-quark events that contain no jets other than the $b$-tagged jet, and $\epsilon_{\textrm{jet-veto}}^\text{MC}$ extrapolates the top-quark MC prediction from the top CR (without $H_{\textrm{T}}$ requirement) to the signal region.
The $\epsilon_\text{single-jet-veto}$ is determined both in data and simulation using events with two leptons, the same requirements on \ptmiss, $\pT^{e\mu}$ and $m_{e\mu}$ as for the signal selection, and at least one $b$-tagged jet. The small contributions to this region of the signal and other background contributions, mainly $W$+jets production, are subtracted before the calculation of $\epsilon_\text{single-jet-veto}^\text{Data}$.
The ratio $\epsilon_\text{single-jet-veto}^\text{Data}/\epsilon_\text{single-jet-veto}^\text{MC}$ then corrects for differences in the veto efficiency for a single jet between data and simulation. It is found to be consistent with one.
The exponent $\langle n_\text{jets} \rangle$ represents the average number of jets in the top CR and is measured to be approximately 2.5 in both data and top-quark background simulation. It is varied by $\pm1.0$ as part of the uncertainty in the method to conservatively cover $\langle n_\text{jets} \rangle$ variations in different control regions as well as variations due to detector uncertainties and modelling, with a small impact (1.8\%) on the total uncertainty in the top-quark background estimate.
 
\begin{figure}[t]
\centering
\includegraphics[width=0.48\textwidth]{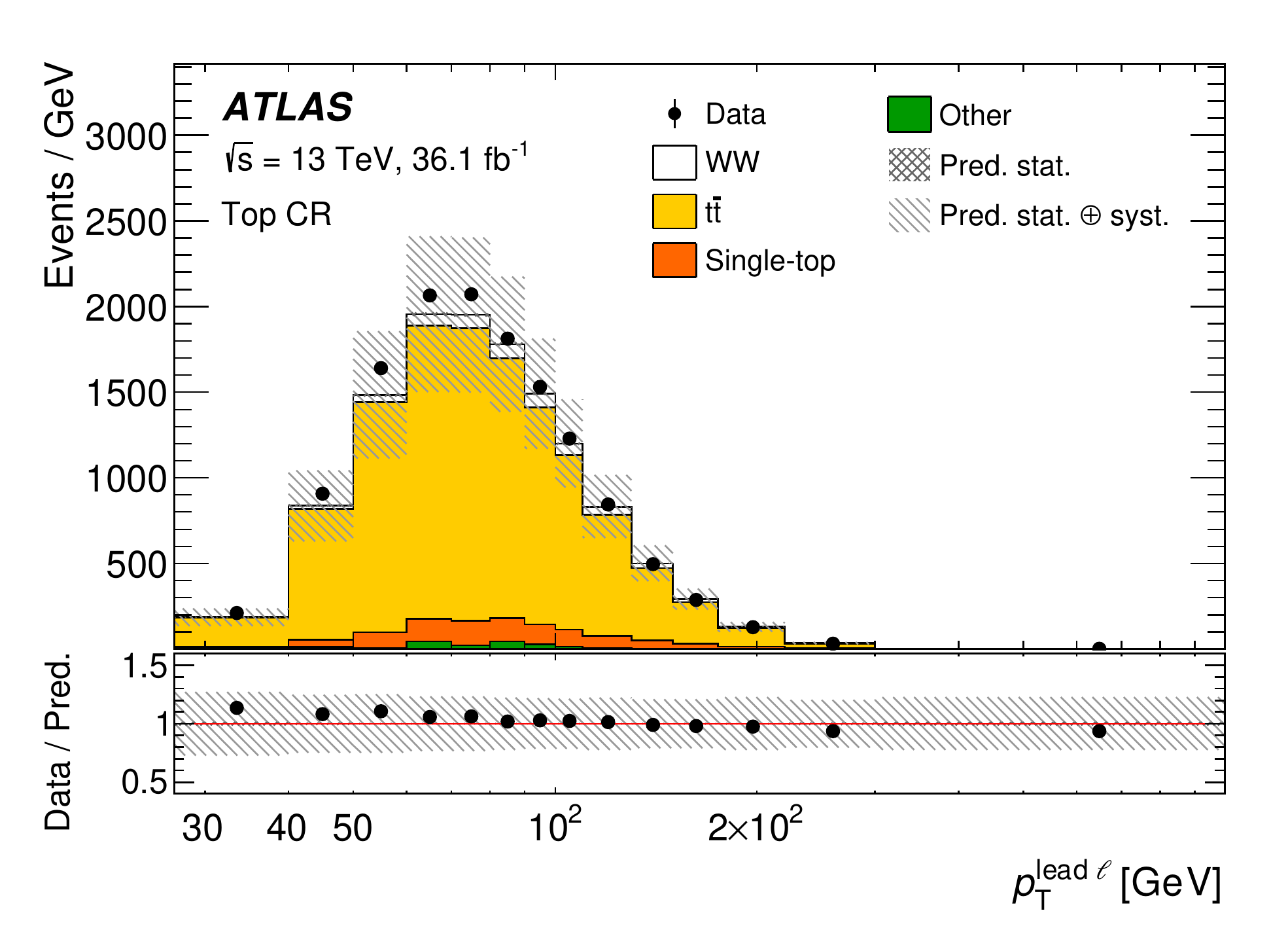}
\includegraphics[width=0.48\textwidth]{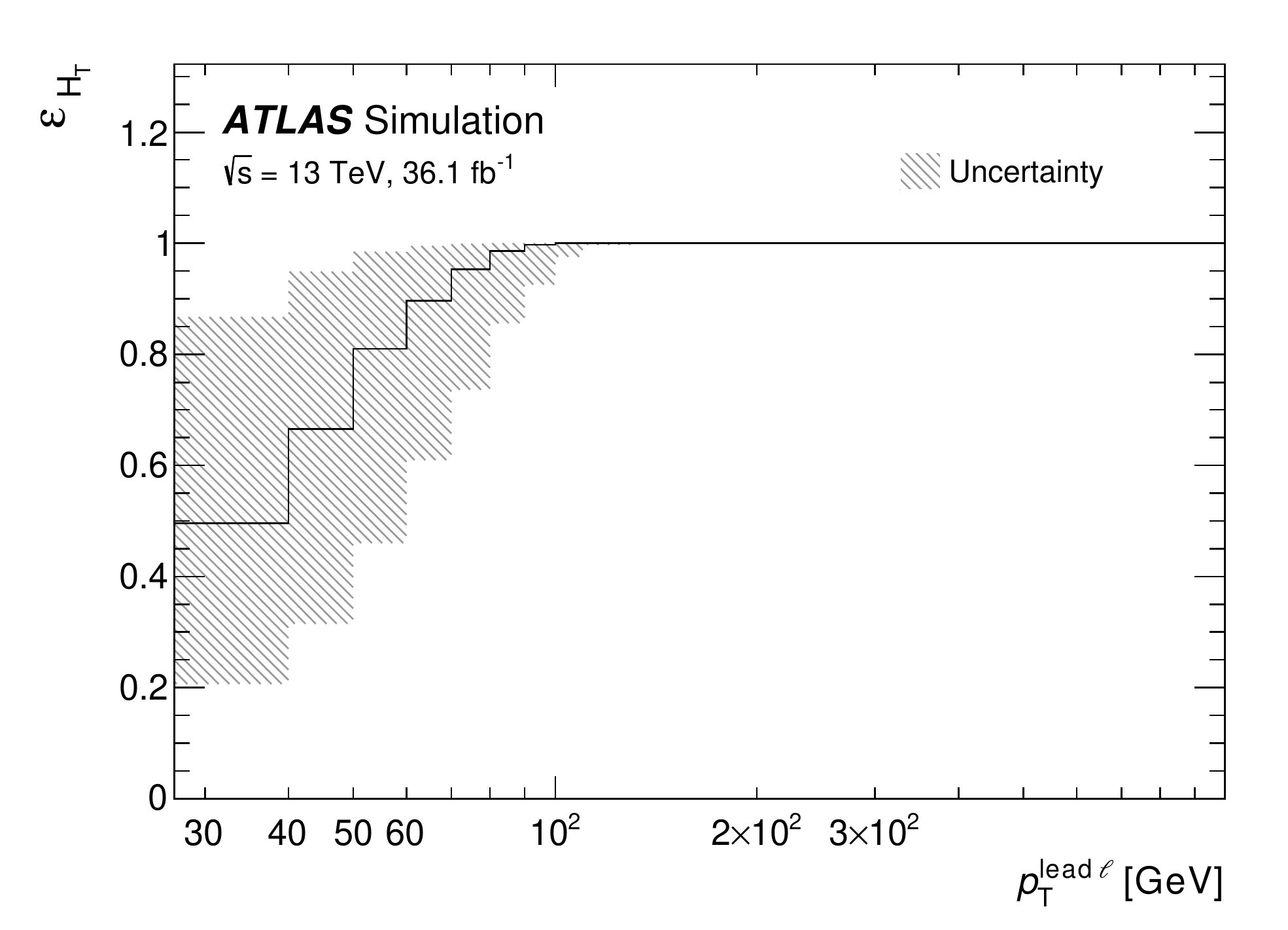}
\includegraphics[width=0.48\textwidth]{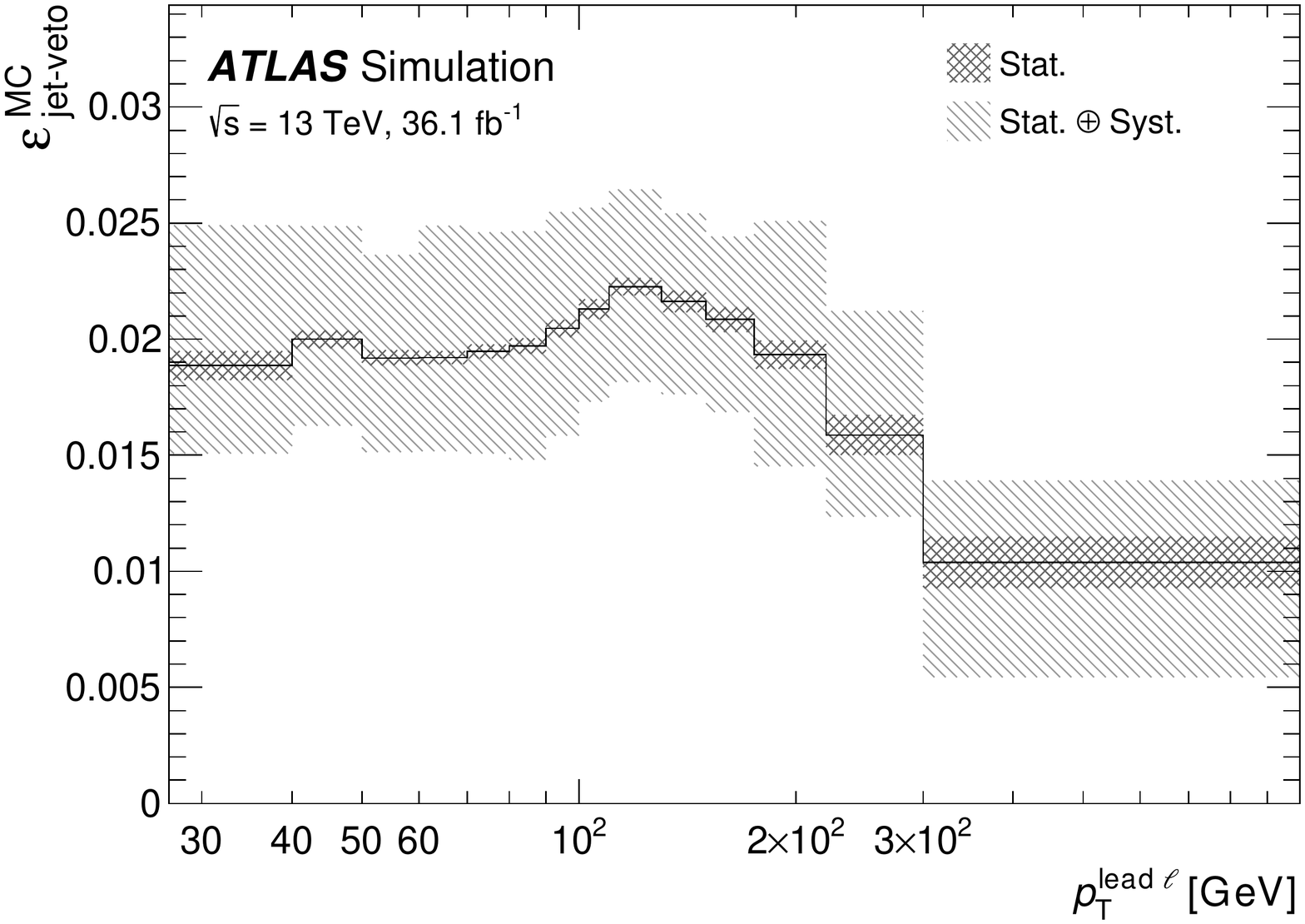}
\includegraphics[width=0.48\textwidth]{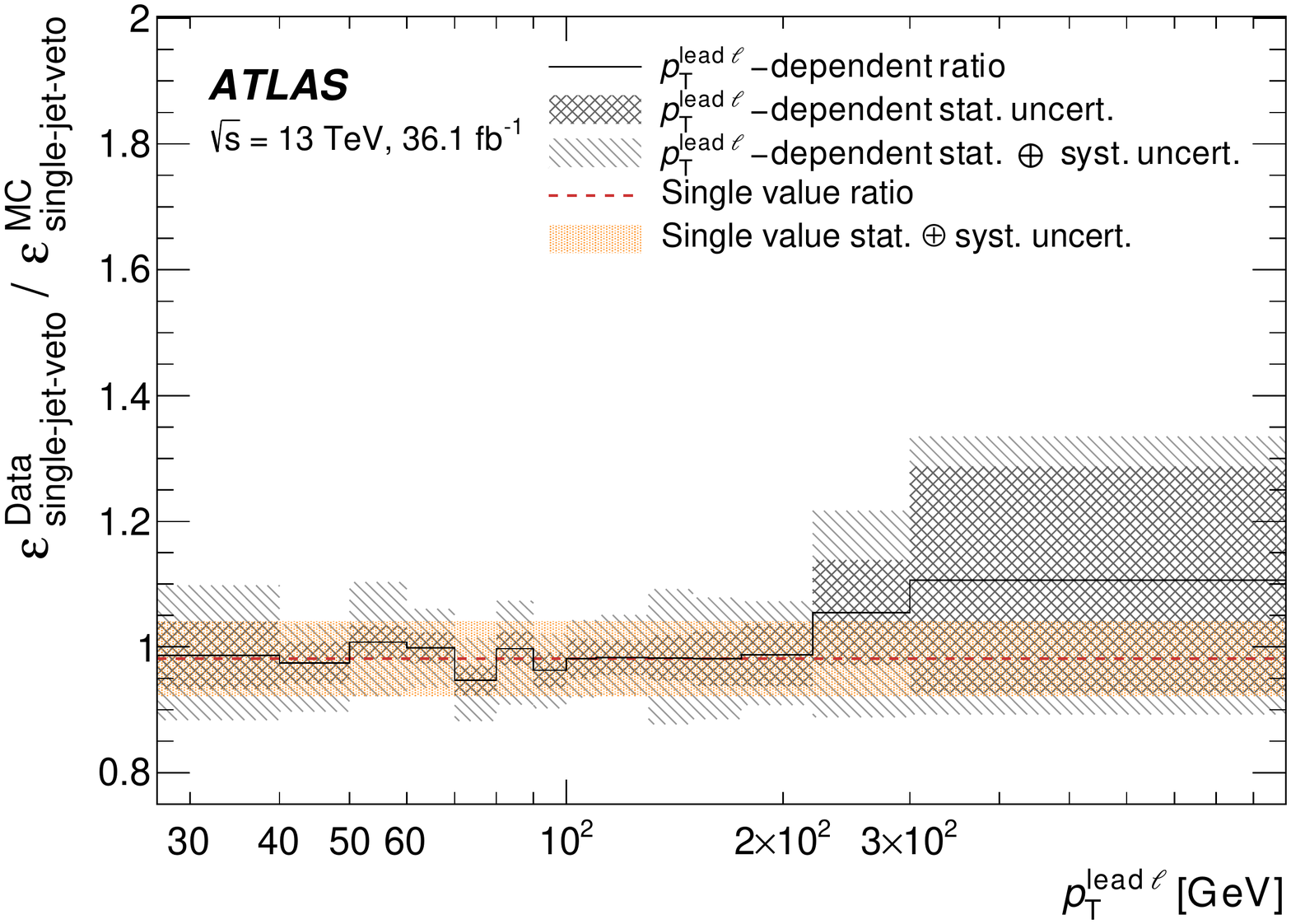}
\caption{
Inputs to the partly data-driven method for the top background estimate as a function of the \pt of the leading lepton: (upper left) events selected in data and in simulation in the top CR, with a requirement of $H_{\textrm{T}}>200\,~\gev$ applied, (upper right) the $H_{\textrm{T}}$ cut efficiency $\epsilon_{H_{\textrm{T}}}$, (lower left) the MC-based jet-veto efficiency $\epsilon_\text{jet-veto}^\text{MC}$, and (lower right) the efficiency ratio $\epsilon_\text{single-jet-veto}^\text{Data} / \epsilon_\text{single-jet-veto}^\text{MC}$. The latter is constant within uncertainties, and therefore replaced by the inclusive efficiency ratio (dashed line). In all figures, statistical and systematic uncertainties are displayed as hatched bands.
}
\label{fig:top_jvsp_body}
\end{figure}

The top-quark background estimate includes detector uncertainties in addition to the uncertainties in the method.
Modelling uncertainties are determined using alternative MC samples and include the modelling of the parton shower, extra QCD radiation and the effect of the choice of generators.
Interference effects between $Wt$ and \ttbar\ are also considered.
These modelling uncertainties are estimated by comparing the results from different MC samples described in Section~\ref{sec:dataMCsamples}.
The cross-section uncertainty is taken to be 6\% for \ttbar~\cite{ttbarXsec1, ttbarXsec2, ttbarXsec3, ttbarXsec4, ttbarXsec5, ttbarXsec6, ttbarXsec7} and 10\% for $Wt$ production~\cite{Kidonakis:2010ux,Kidonakis:2013zqa}.
The total uncertainty in the top-quark background estimate in the signal region is about 12\% using this partly data-driven approach, making use of cancellations of systematic uncertainties in the ratio $\epsilon_{\textrm{jet-veto}}^\text{MC}/(\epsilon_\text{single-jet-veto}^\text{MC})^{\langle n_\text{jets} \rangle}$ in Eq.~(\ref{epsilon_jet_veto}). It is dominated by the $b$-tagging and modelling uncertainties. The contribution of the $t\bar{t}$ and $Wt$ background to the total expected yield in the signal region is about 25\% (17\% $t\bar{t}$ and 8\% $Wt$).
 
The differential top-quark background contribution and its uncertainties are evaluated by applying the same procedure in each bin of the measured observables. As an example, \figref{top_jvsp_body} shows the relevant quantities used in this partly data-driven method, as a function of the transverse momentum of the leading (highest $\pT$) lepton. The systematic uncertainties in $N^{\textrm{top}}_{\textrm{SR}}$ are significantly reduced due to the systematics cancellations compared with the uncertainty bands from \figref{top_jvsp_body}.
The decrease of $\epsilon_\text{jet-veto}^\text{MC}$ at high leading lepton \pt is due to an increase in the typical \pt of extra jets which recoil against the leptons, nearing the jet-veto \pt threshold, and hence reducing the probability to still pass the jet veto.
Since the efficiency ratio, $\epsilon_\text{single-jet-veto}^\text{Data} / \epsilon_\text{single-jet-veto}^\text{MC}$, is found to be independent of any kinematic variable, the single value of $0.98 \pm 0.05$ is used for all differential distributions. This is shown as a dashed line in the lower right panel of \figref{top_jvsp_body}.

\subsection{Background from Drell--Yan production}
 
The estimate of the Drell--Yan background process is based on MC simulation, with a $5\%$ theoretical cross-section uncertainty~\cite{STDM-2015-03}.
A validation region dominated by Drell–Yan events is defined with the same selections as for the signal region, but with the $e\mu$ invariant mass required to be $45~\GeV<m_{e\mu}<80$~\GeV\, and with the events failing either the $\pT^{e\mu}$- or the \ptmiss -requirement to make the sample orthogonal to that in the signal region.
Good agreement between the data and the simulation is observed in this region.
The shape uncertainty is evaluated by using an alternative MC event generator, as detailed in \secref{dataMCsamples}, and includes uncertainties due to the modelling of the acceptance.
The total uncertainty in the Drell--Yan background is 11\% and
the contribution of this background in the signal region is found to be 4\%.

\subsection{Background from $W$+jets production}
 
The yield of $W$+jets is estimated by comparing in data the number of events with leptons satisfying either of two alternative sets of selection requirements, together with the \WW signal selection criteria, following the same procedure as that described in Ref.~\cite{STDM-2013-07}.
The {\it loose} lepton selection criteria are defined such that the signal sample is a subset of the loose lepton sample.
For electrons, the loose selection corresponds to the {\it MediumLH} quality definition~\cite{Aaboud:2019ynx} and no isolation requirements are imposed.
For muons, the loose selection is the same as for signal muons, except that the isolation requirement is omitted.
The {\it tight} selection criteria are the same as those used for the signal selection.
With the introduction of real-lepton and fake-lepton efficiencies, a system of four equations can be solved to estimate the number of $W$+jets events.
Here, the number of events that have exactly one loose muon (electron) and one tight electron (muon), two loose leptons or two tight leptons, are used.
The real-lepton (fake-lepton) efficiency used in these equations is defined as the probability for prompt (fake) leptons selected with the loose criteria to satisfy the tight selection criteria.

The efficiencies for real electrons (muons) are determined using MC simulation, with data-to-MC correction factors~\cite{Aaboud:2019ynx,PERF-2015-10} applied. The efficiencies for fake electrons (muons) are measured using a multijet data sample, in a control region with exactly one loose electron (muon) and between one and three jets.
 
Events in this control region are also required to have low \ptmiss and low transverse mass\footnote{The transverse mass is defined as: $\mt=\sqrt{2\pt^\ell\ptmiss\left(1-\cos{\left(\Delta\phi(\ell,\ptmiss)\right)}\right)}$.} \mt, to fulfil angular requirements between \ptmiss and the jets in the event, and to have no $b$-tagged jets.
Real-lepton contributions to the control region are estimated using MC simulation and are subtracted.
 
Both the real- and fake-lepton efficiencies are derived as functions of \pT\ and $\eta$ of the lepton. This is sufficient to describe the most important correlations with the differential distributions studied.
Moreover, as the loose lepton selection in the $W$+jets background estimate at low lepton-\pt ($\pt<50$ \GeV\ for muons or $\pt<60$ \GeV\ for electrons) is typically looser than in the trigger selection, the efficiencies are provided separately for low-\pt electrons or  muons that satisfy or fail to satisfy the trigger selection requirements.
The fake-lepton efficiency for the non-triggered leptons is estimated using events recorded with triggers that have lower muon-\pt, only \emph{MediumLH} electron quality and no lepton isolation requirements, but only record a fraction of the events satisfying these criteria.
 
The uncertainty in the \Wjets\ background is directly related to the uncertainties in the real- and fake-lepton efficiencies.
For real-lepton efficiencies, these take into account uncertainties in electron and muon reconstruction and isolation correction factors.
Uncertainties in the fake lepton efficiencies include variations in the control region definition, as well as normalization and shape uncertainties in the subtracted contributions from other processes in the control region.
The control region variations are designed to cover the uncertainty in the flavour composition of the jets faking leptons, and include variations of the \mt\ requirement and the number of $b$-tagged jets.
 
The total uncertainty in the $W$+jets yield is 90\% and is dominated by the uncertainty in the fake  and real electron efficiencies, because of the greater contribution of electron fakes to the $W$+jets background.
The $W$+jets background amounts to 3\% of the expected yield in the signal region.
 
The differential $W$+jets distributions necessary for the differential cross-section measurements are also determined in a fully data-driven way, by evaluating the same system of linear equations~\cite{STDM-2013-07} in each bin of the differential distributions.
 
\begin{figure}[t!]
\includegraphics[width=.47\textwidth]{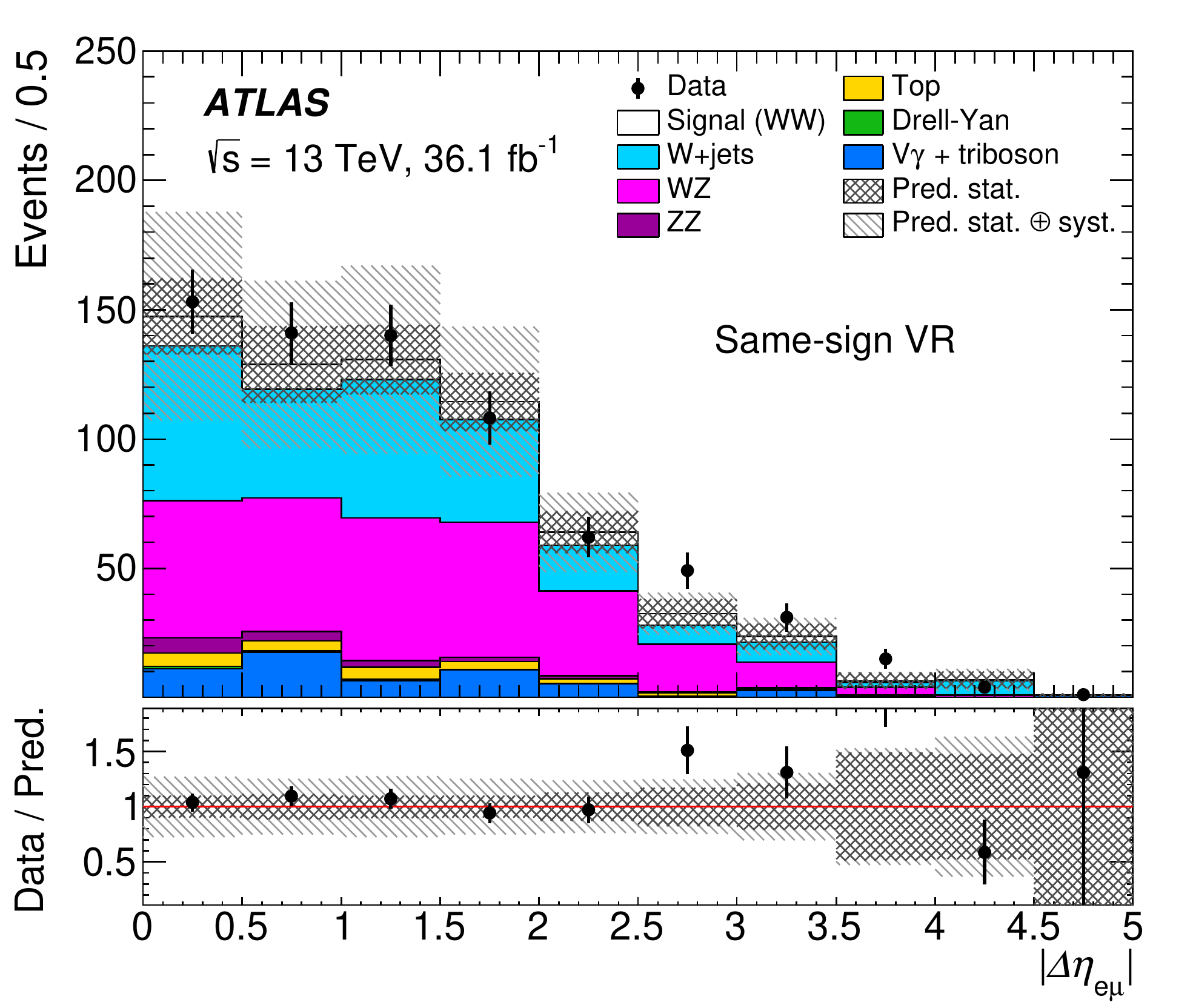}
\includegraphics[width=.47\textwidth]{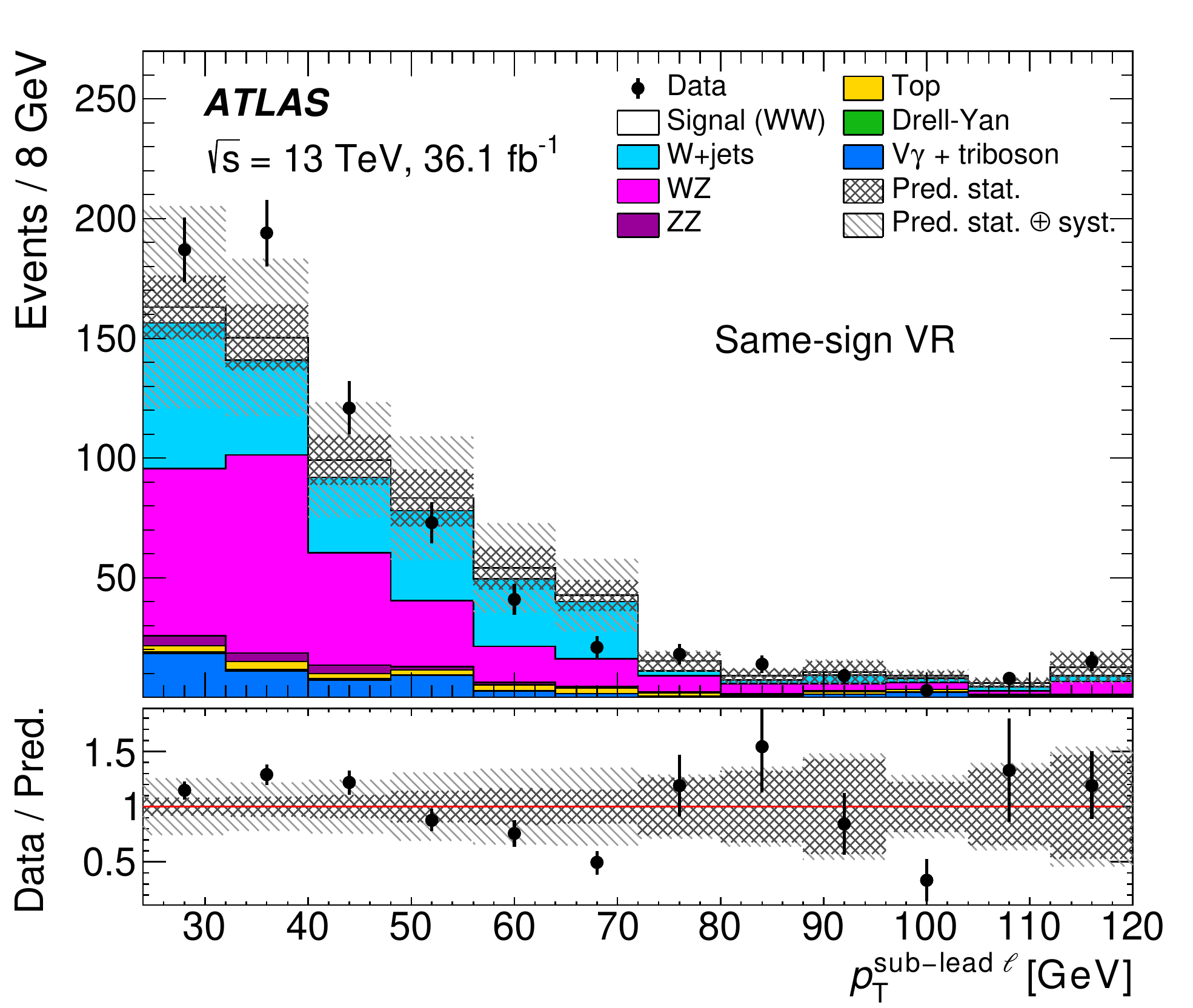}
\caption{Distributions of the pseudorapidity difference between the leptons (left) and the transverse
momentum of the sub-leading lepton (right) for the same-sign validation region. The uncertainties shown include statistical and systematic uncertainties.}
\label{fig:ValidationRegions}
\end{figure}

The predicted contributions to the backgrounds from $W$+jets are validated
using a data control sample in which the two selected leptons are required to have the same electric charge
(same-sign) and satisfy all the other selection requirements.
Figure~\ref{fig:ValidationRegions} shows the pseudorapidity difference between the leptons and the transverse momentum of the sub-leading lepton for this same-sign control sample. The predictions and the data agree well.
 
\subsection{Background from multi-boson production}
 
The estimate of the diboson background from $WZ$, $ZZ$, $W\gamma$ and $Z\gamma$ processes is based on MC simulation.
These processes contribute about 3\% to the total number of events.
The uncertainty in the cross-section for these diboson processes is taken as 10\%~\cite{ATL-PHYS-PUB-2016-002,Grazzini:2015nwa} and variations in the shape and the acceptance are considered for $WZ$ and $ZZ$ production by using alternative MC generators, as detailed in~\secref{dataMCsamples}.
 
The $V\gamma$ background simulation is validated in data using the events passing the same selection as for the signal region, except inverting the electron identification criteria and requiring the reconstructed electron track to have no hit in the innermost layer of the pixel detector.
The $WZ$ background simulation is validated in data using events that allow for the presence of a third loosely isolated lepton with $\pt>10$~\gev\ and require the same-flavour lepton pair to be of opposite sign and with invariant mass of $80~\gev<m_{ee/\mu\mu}<100~\gev$, while otherwise passing the signal region selection.
Good agreement between the data and the simulation is found in both regions.
 
The background from triboson production ($WWW$, $WWZ$, $WZZ$ and $ZZZ$) is less than 0.1\% and is evaluated using MC simulation.
The cross-section uncertainty is taken as $30\%$~\cite{ATL-PHYS-PUB-2016-002}.
 
\subsection{$WW$ candidate events and estimated background yields}\label{sec:yieldtable}
 
\begin{figure}[h!]
\centering
\includegraphics[width=0.46\textwidth]{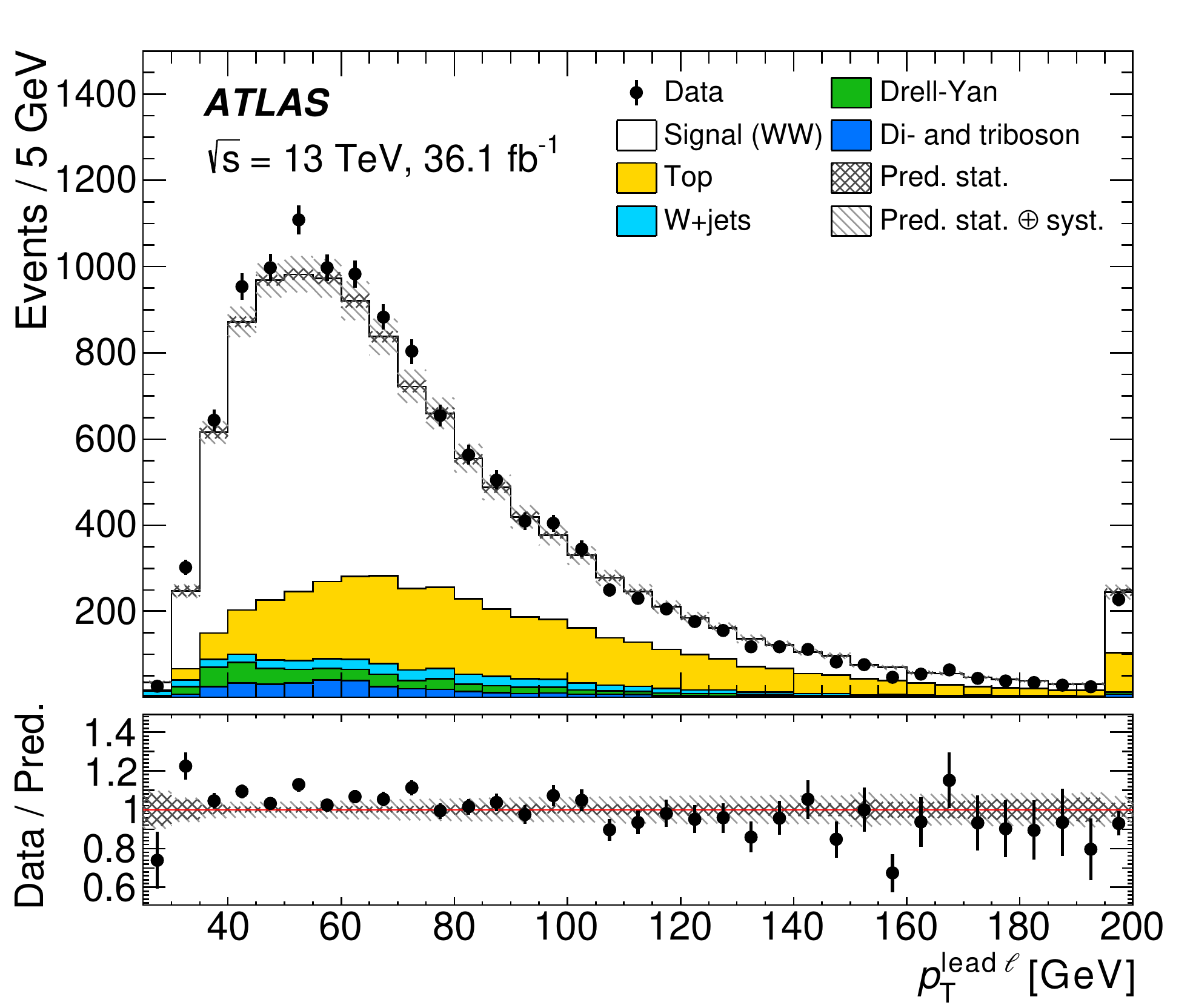}
\includegraphics[width=0.46\textwidth]{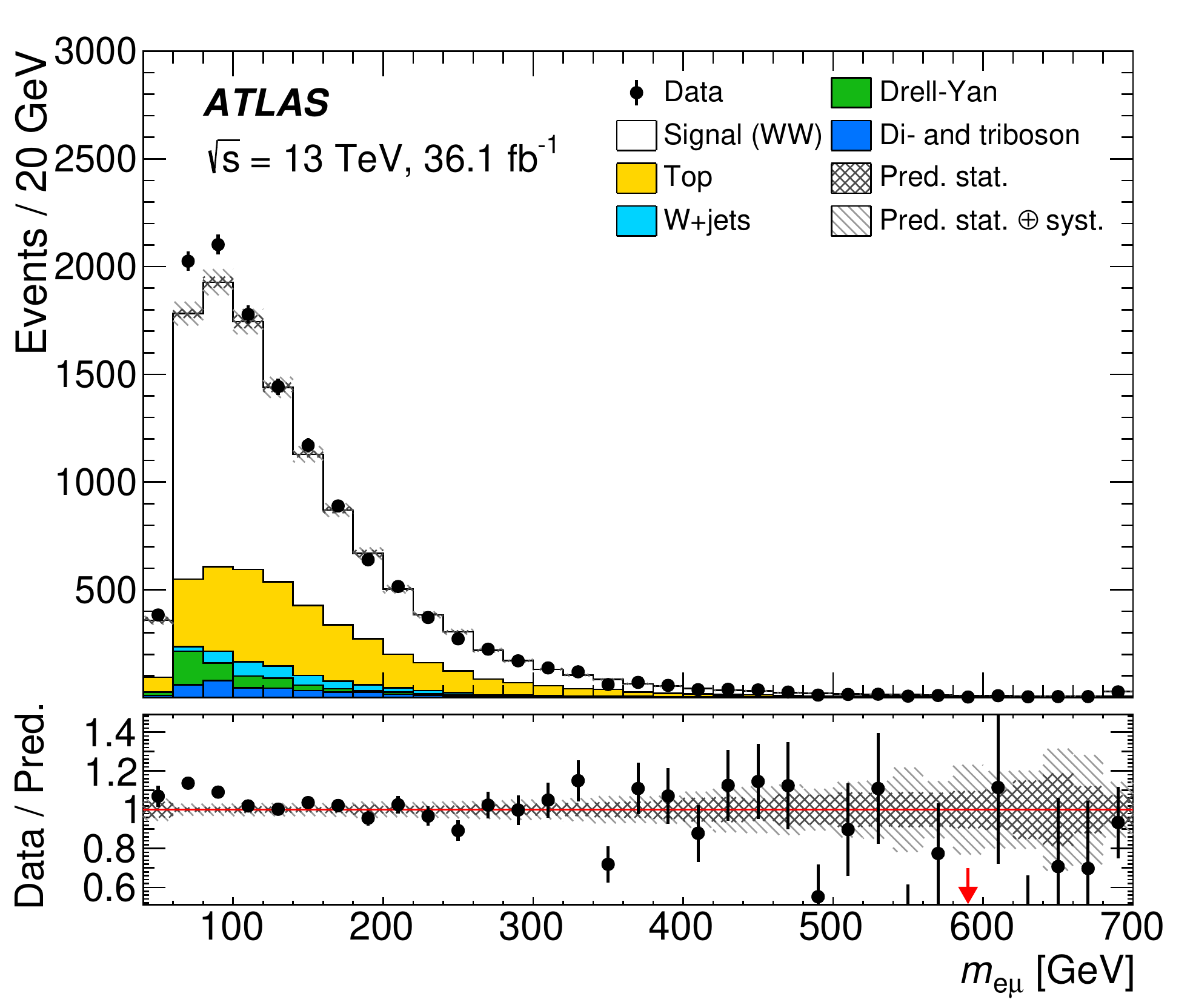}
 
\includegraphics[width=0.46\textwidth]{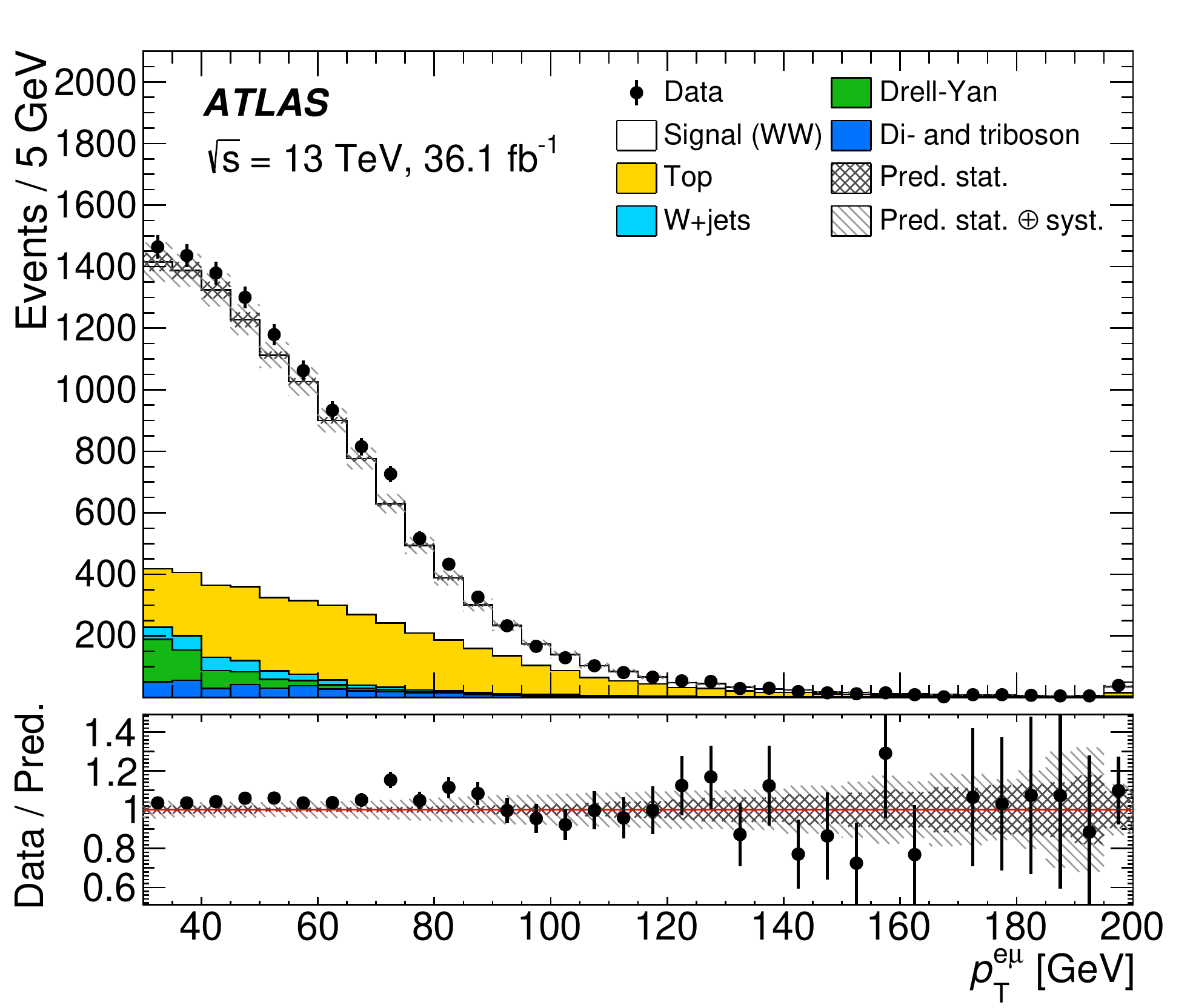}
\includegraphics[width=0.46\textwidth]{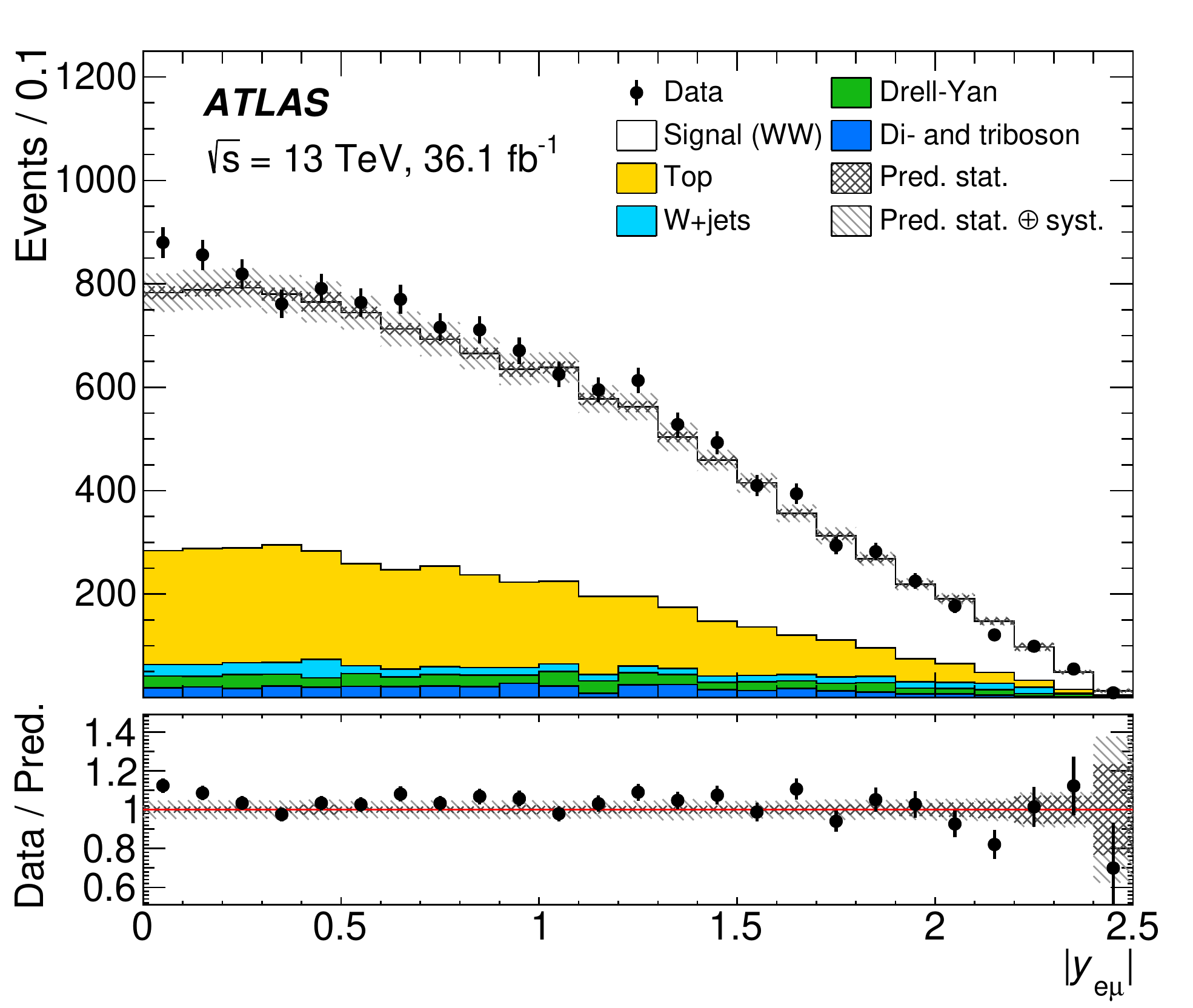}
 
\includegraphics[width=0.46\textwidth]{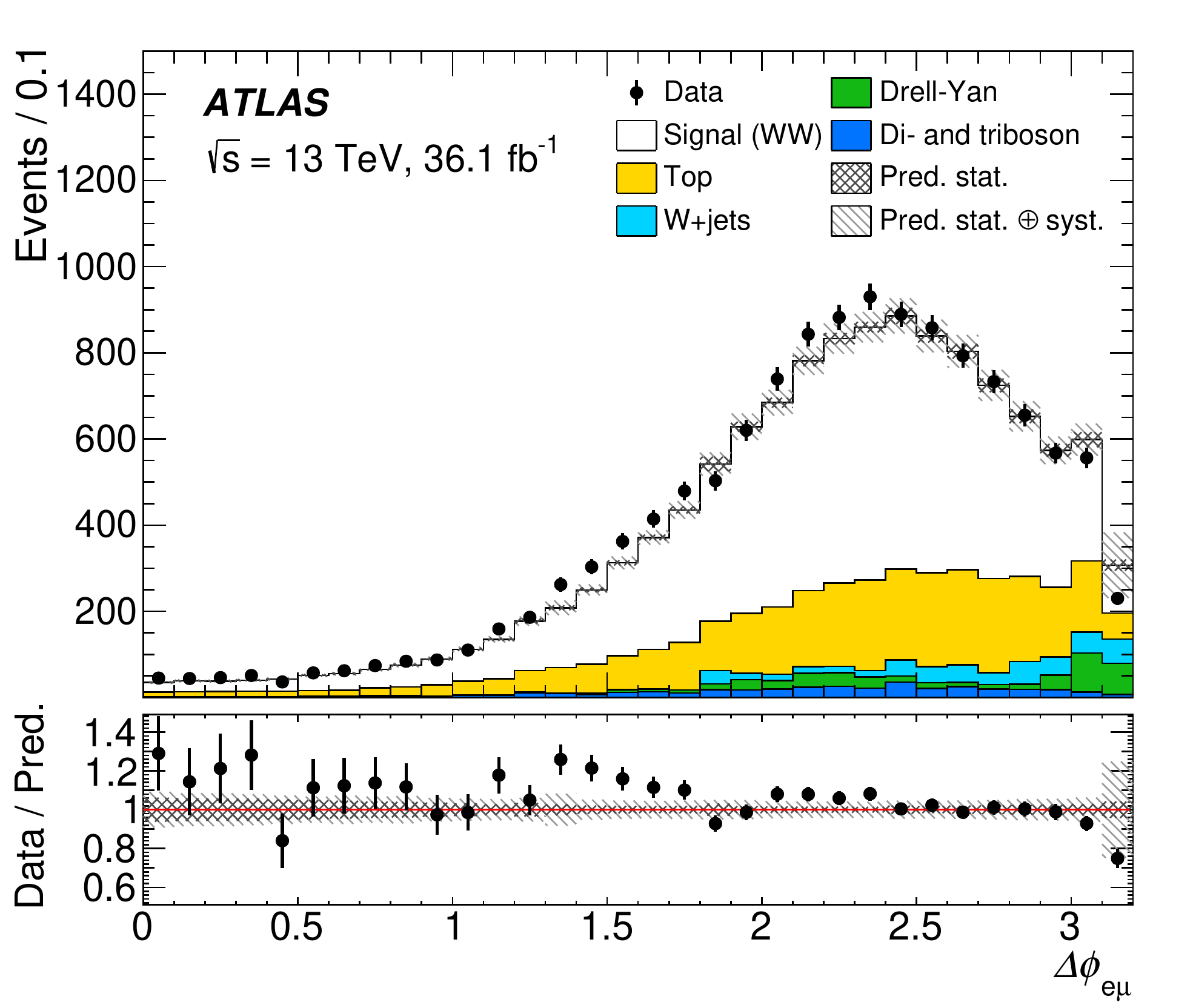}
\includegraphics[width=0.46\textwidth]{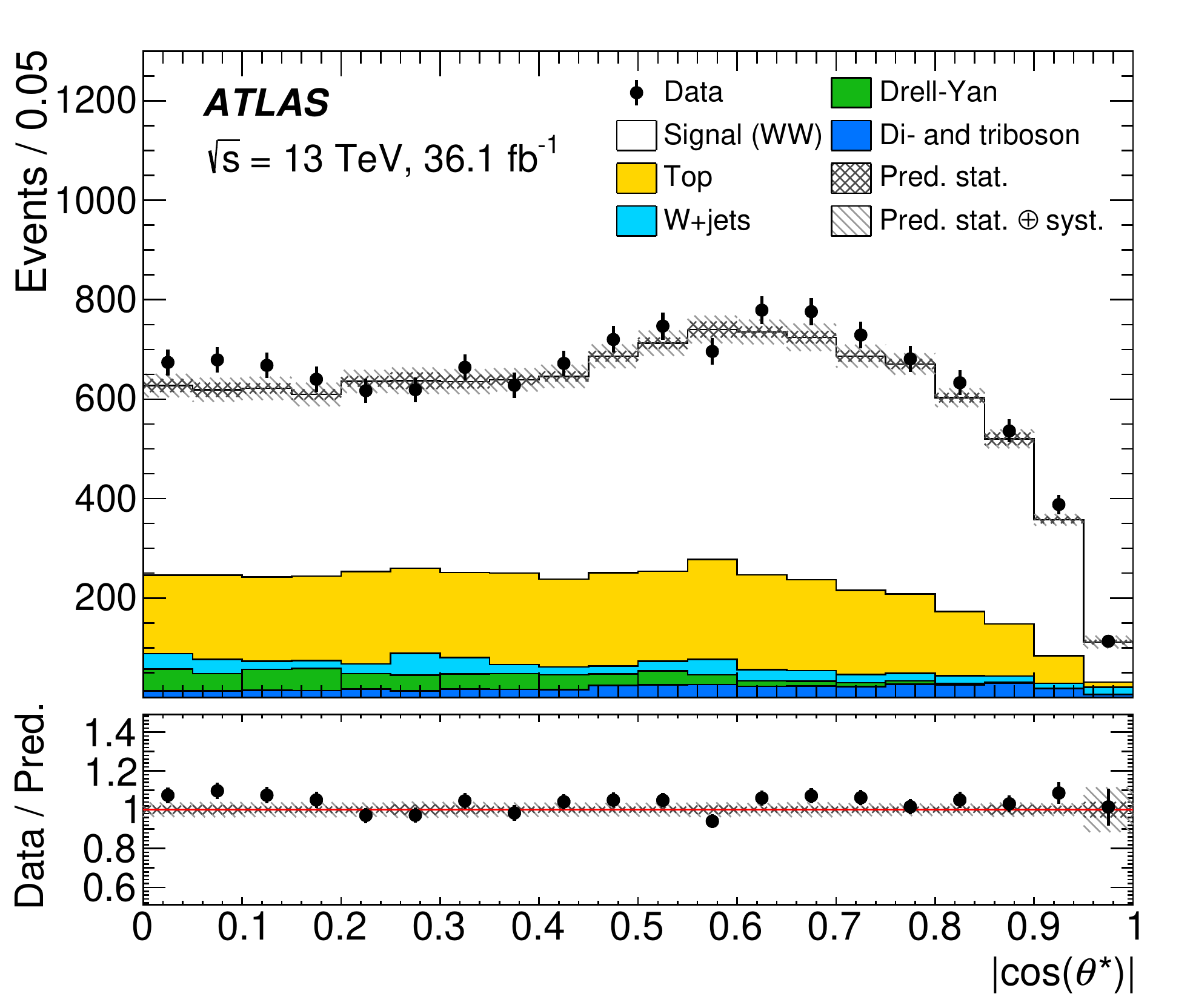}
\caption{Kinematic distributions of the selected data events after the full event selection (from left to right and top to bottom):  \ptlzero, \mll, \ptll, \yll, \dphill and \costheta.
Data are shown together with the predictions of the signal and background production processes.
Statistical and systematic uncertainties in the predictions are shown as hatched bands.
The lower panels show the ratio of the data to the total prediction. An arrow indicates that the point is off-scale.
The last bin includes the overflow.
}
\label{fig:sr_control_plots}
\end{figure}

After applying all the selection requirements, 12\ 659 events are observed in data, with a contribution of 65\% from \WW production, which is estimated using simulation (see Section~\ref{sec:dataMCsamples}).
A summary of the data, signal, and background yields is shown in Table~\ref{tab:Postfit_EventsTable}.
Kinematic distributions comparing the selected data with the signal and backgrounds in the signal region are shown in Figure~\ref{fig:sr_control_plots}.
Fair agreement between data and expectations is observed for the overall normalization and the shapes of various kinematic distributions.
Small underpredictions in the peak region of the leading lepton \pt distribution, the low \mll region and a small downward trend in the ratio of the data to expectations in the \dphill distribution have also been observed in the previous ATLAS measurement at $\sqrt{s}=8$~\TeV~\cite{STDM-2013-07}. The trend in the \dphill distribution was also observed at $\sqrt{s}=8$~\TeV~by CMS Collaboration~\cite{Khachatryan:2015sga}.
For the  \dphill distribution, the largest discrepancy between data and expectations (of about three standard deviations) is observed in the range $1.3 <  \dphill < 1.6$.
 
\begin{table}[t!]
\caption{Number of events observed in data, compared with the numbers of predicted signal and background events in the signal region.
The systematic uncertainties, described in Section~\ref{sec:systematics}, do not include the uncertainty in the integrated luminosity.
The uncertainties in the total background and in the sum of signal and background are the sums in quadrature of the uncertainties in the various background and signal sources.}
\centering
\begin{tabular}{ l  S[table-format=5,table-number-alignment=right]@{$\,\,$~~~~~~~~~~~}  S@{~~~~~~~~~~~~~$\,\pm\,$}S[table-format=2,table-number-alignment=left]  S@{~~~~~~~~~~~~$\,\pm\,$}S[table-format=3,table-number-alignment=left]}
\toprule
& \multicolumn{1}{c}{Number of events} & \multicolumn{2}{c}{Statistical uncertainty}  & \multicolumn{2}{c}{Systematic uncertainty}\\
\midrule
Top-quark & 3120 & & 50  & & 370 \\
Drell--Yan & 431 & & 13 & & 44 \\
$W$+jets & 310 & & 60  & & 280 \\
$WZ$ & 290 & & 11 & & 33 \\
$ZZ$ &  16 & & 1 & & 2 \\
$V\gamma$ &  66 & & 11 & & 10 \\
Triboson & 8 & & 1 & & 3 \\
Total background & 4240 & & 80 & & 470 \\
\hline
Signal $(WW)$ & 7690 & & 30  & & 220 \\
Total signal+background & 11930 & & 90 & & 520 \\
\hline
Data & 12659 & \multicolumn{2}{c}{-} & \multicolumn{2}{c}{-}  \\
\bottomrule
\end{tabular}
\label{tab:Postfit_EventsTable}
\end{table}

\section{Fiducial cross-section determination}
\label{sec:crosssection}
 
The \WW cross-section is evaluated in the fiducial phase space of the $e\mu$ decay channel, as defined in Table~\ref{tab:fiducialtruthlevelcuts}.
In simulated events, electrons and muons are required to originate from one of the $W$ bosons produced in the hard scatter,
and the momenta of photons emitted in a cone $\Delta R = $ 0.1 around the lepton direction are added to the lepton momentum after QED final-state radiation to form `dressed' leptons~\cite{Buckley:2010ar}.
Final-state particles with lifetimes greater than 30~ps are clustered into jets (referred to as particle-level jets) using the same algorithm as for detector-level jets, i.e.\ the anti-$k_t$ algorithm with radius parameter $R$ = 0.4.
The selected charged leptons and any neutrino in the event are not included in the jet clustering.
The fiducial phase space at particle level does not make any requirement on jets containing $b$-quarks.
The missing transverse momentum is defined at particle level as the transverse component of the vectorial sum of the neutrino momenta. Its magnitude is denoted in Table~\ref{tab:fiducialtruthlevelcuts} by $\met$.
 
The fiducial cross-section is obtained as follows:
\begin{linenomath*}
\begin{equation*}
\sigma^{\mathrm{fid}} _{\WWemu}
= \frac{N_{\rm obs} - N_{\rm bkg}}{C\, \times {\cal L} },
\label{eq:xSectFid}
\end{equation*}
\end{linenomath*}
where ${\cal L}$ is the integrated luminosity, $N_{\rm obs}$ is the observed number of events, $N_{\rm bkg}$ is the estimated number of background events and $C$ is a factor that accounts for detector inefficiencies, resolution effects and contributions from $\tau$-lepton decays.
The $C$ factor is defined as the ratio of the number of reconstructed $WW$ events after the final selection with electrons or muons in the final state (including electrons or muons from $\tau$-lepton decays) to the number of $WW$ events generated in the fiducial region where only direct decays of $W$ bosons to electrons and muons are allowed.
The $C$ factor takes into account the contribution to the \WW signal originating outside of the fiducial phase space.
This contribution is estimated from MC simulation to be about 21\% of the expected reconstructed signal, about 40\% of which originates from $\tau$-lepton decays.
The $C$ factor has a value of 0.613 with an uncertainty of 3\%, including experimental and unfolding method sources, as detailed in \secref{systematics}.
 
The fiducial cross-section as a function of the jet-veto \pt threshold is determined using the same method, but modifying the
selection requirements
to exclude events with jets above a transverse momentum of 30~\gev, 35~\gev, 40~\gev, 45~\gev, 50~\gev, 55~\gev, and 60~\gev, respectively. The values for $C$ are determined for each threshold and increase from 0.598 to 0.625.
 
The differential cross-sections are determined using an iterative Bayesian unfolding method~\cite{BayesUnfold1,BayesUnfold2} with one iteration for \mll, \ptlzero, \yll, \dphill and \costheta, and two iterations for \ptll.
The number of iterations is optimized to find a balance between too many iterations, causing high statistical uncertainties in the unfolded distributions, and too few iterations, which can bias the measurement towards the MC prediction.
The unfolding procedure corrects for migrations between bins in the distributions during the reconstruction of the events, and applies fiducial as well as reconstruction efficiency corrections. The fiducial corrections take into account events that are reconstructed in the signal region, but originate from outside the fiducial region; the reconstruction efficiency corrects for events inside the fiducial region that are not reconstructed in the signal region due to detector inefficiencies. Tests with MC simulation demonstrate that the method is successful in retrieving the true distribution in the fiducial region from the reconstructed distribution in the signal region.
 
\begin{table}[t!]
\caption{Definition of the $WW\rightarrow e\mu$ fiducial phase space. }
\begin{center}
\begin{tabular}{rp{0.1cm}l}
\toprule
\multicolumn{3}{c}{Fiducial selection requirements}   \\
\midrule
$\pt^{\ell}$ & > & $27$~\GeV  \\
$|\eta^{\ell}|$ & < & $2.5$         \\
$\mll$ & > & $55$~\GeV               \\
$\ptll$ & > & $30$~\GeV \\
$\met$ & > & $20$~\GeV \\
No jets with $\pt$ & > & $35$ \GeV, $|\eta| <$ 4.5 \\
\bottomrule
\end{tabular}
\end{center}
\label{tab:fiducialtruthlevelcuts}
\end{table}

 
\section{Systematic uncertainties}
\label{sec:systematics}
 
Systematic uncertainties in the $WW$ cross-section measurements arise from the reconstruction of leptons and jets, the background determination, pile-up and integrated luminosity uncertainties, as well as the procedures used to correct for detector effects, and theoretical uncertainties in the signal modelling.
 
For leptons and jets, uncertainties in the momentum or energy scale and resolution are considered~\cite{PERF-2015-10, PERF-2017-03, PERF-2016-04}.
Uncertainties in the lepton reconstruction and identification efficiencies~\cite{Aaboud:2019ynx, PERF-2015-10} as well as the efficiency of the jet vertex tagging requirements~\cite{PERF-2014-03,PERF-2016-06} in the simulation are taken into account.
Uncertainties in the $b$-tagging, which mainly stem from the top-quark background contributions, are also taken into account based on the studies in Refs.~\cite{btagEffs,ATLAS-CONF-2018-001}.
The impact of uncertainties in the scale and resolution of \ptmiss\ are estimated as discussed in Ref.~\cite{PERF-2016-07}.
The pile-up modelling uncertainty is evaluated by varying the number of simulated pile-up interactions by its uncertainty of 10\% of the nominal value.
The variations are designed to cover the uncertainty in the ratio of the predicted to the measured cross-section of non-diffractive inelastic events producing a hadronic system of mass $m_X > 13~\GeV$~\cite{STDM-2015-05}, where the nominal value of $\sigma_{\textrm{inel}}=74$ mb is used in the simulation.

Uncertainties in MC-based background processes include variations of the shapes of predicted distributions, the normalization, and the statistical uncertainties in the simulation, in addition to the full set of detector-related uncertainties. The first two are estimated as discussed in Sections~\ref{sec:dataMCsamples} and~\ref{sec:backgrounds}.
The uncertainties in the background from top-quark and $W$+jets production are mitigated by the use of the data-driven methods described in Section~\ref{sec:backgrounds}.

Uncertainties due to the unfolding procedure and the modelling of the signal process are considered by repeating the cross-section extraction with modified inputs.
The uncertainty due to the choice of generator for the hard interaction, the parton shower model and the underlying-event modelling for the MC-based unfolding inputs, is estimated by using \SHERPAV{2.2.2} instead of \POWHEGBOX{}+\PYTHIAV{8} for \qqbar-initiated $WW$ production, with the samples detailed in Section~\ref{sec:dataMCsamples}.
The impact of mismodelling of the data by \POWHEGBOX{}+\PYTHIAV{8} for each observable is estimated by reweighting the distribution at generator level to improve the agreement between data and simulation after event reconstruction. The obtained prediction at detector level, which is then very similar to data, is unfolded with the normal inputs and the difference from the reweighted prediction at generator level is considered as an uncertainty.
The impact of statistical uncertainties in the nominal signal simulation is estimated using pseudo-data.
The theory uncertainties cover PDF and scale variations of the unfolding inputs. The PDF uncertainty is estimated as the 68\% confidence level (CL) envelope of the \ctten~\cite{CTEQ10} prediction.
The uncertainty associated with higher-order QCD corrections is evaluated by varying the renormalization (\muR) and factorization (\muF) scales independently by factors of 2 and 0.5 with the constraint $0.5\leq\muF/\muR\leq2$.

The uncertainty in the combined 2015+2016 integrated luminosity is 2.1\%. It is derived from the calibration of the luminosity scale using $x$-$y$ beam-separation scans, following a methodology similar to that detailed in Ref.~\cite{DAPR-2013-01}, and using the LUCID-2 detector for the baseline luminosity measurements \cite{LUCID2}.
The LHC beam energy uncertainty is estimated to be 0.1\%~\cite{Todesco:2017nnk}. It affects the signal cross-section by less than 0.2\% and is not considered in the total uncertainty.
 
A summary of the systematic uncertainties in the fiducial cross-section measurement is shown in \tabref{systs}.
The total uncertainty is dominated by the $b$-tagging uncertainty (3.4\%), the jet energy scale uncertainty (3\%), and the modelling of the $W$+jets (3.1\%) and top-quark (2.6\%) backgrounds.

\begin{table}[t!]
\caption{Relative uncertainties in the \WW\ fiducial cross-section measurement.}
\centering
\begin{tabular}{lc}
\toprule
Uncertainty source  & Uncertainty [\%] \\
\midrule
Electron & 0.7 \\
Muon & 0.9 \\
Jets & 3.0 \\
$b$-tagging & 3.4 \\
\ptmiss & 0.4 \\
Pile-up & 1.6 \\
\Wjets background modelling & 3.1 \\
Top-quark background modelling & 2.6 \\
Other background modelling & 1.3 \\
Unfolding, incl. signal MC stat. uncertainty & 1.4 \\
PDF+scale & 0.1 \\
\hline
Systematic uncertainty & 6.7 \\
Statistical uncertainty & 1.3 \\
Luminosity uncertainty & 2.1 \\
\hline
Total uncertainty & 7.1 \\
\bottomrule
\end{tabular}
\label{tab:systs}
\end{table}
 
\section{Theoretical predictions}
\label{sec:theory}
 
Theoretical predictions are calculated for the fiducial and the differential cross-sections and include the \qqtoWW and $gg\rightarrow WW$ sub-processes.
The $q\bar{q}$-initiated  production makes up 95\% of the total cross-section, while the non-resonant and resonant $gg$-initiated sub-processes account for 5\%.
 
NNLO predictions for the \qqtoWW production cross-sections are determined using the MATRIX program~\cite{Grazzini:2017mhc,Cascioli:2011va,Gehrmann:2015ora},
including off-shell effects and the non-resonant and resonant gluon-initiated contributions at LO. For improved precision, the MATRIX prediction for \qqbar-initiated production is also complemented with NLO corrections to gluon-induced $WW$ production~\cite{Caola:2016trd} and with extra NLO EW corrections that also include the photon-induced ($\gamma\gamma\rightarrow\WW$) contribution~\cite{Biedermann:2016guo}. For all these predictions, the NNPDF 3.1 LUXqed PDF set is used~\cite{Ball:2017nwa,Bertone:2017bme}, the renormalization and factorization scales are set to $m_{WW}/2$, and the scale uncertainties are evaluated according to Ref.~\cite{Stewart:2011cf}.
The PDF uncertainty corresponds to the 68\% CL variations of the NNPDF set.
The MATRIX prediction itself does not include EW radiative effects from leptons in contrast to the MC simulation used to define leptons in the fiducial region, where photons from the parton shower outside a cone of $\Delta R=0.1$ can be present.
The application of NLO EW corrections compensates, at least partially, for this difference.
It is observed that the NLO corrections to the $gg\rightarrow WW$ sub-process increase the fiducial cross-section by 3\%, whereas the NLO EW corrections, applied to the sum of \qq- and $gg$-initiated production, decrease it by 6\%.
 
NLO predictions for \qqtoWW production, which are matched to a parton shower (\qq NLO+PS), are determined using either \POWHEGBOX interfaced to \PYTHIAV{8} or \herwigpp, or \sherpa~2.2.2.
They are combined with the \SHERPA{}+\textsc{OpenLoops} calculation for the gluon-induced non-resonant and resonant \WW production ($gg$ LO+PS).
These predictions are described in detail in \secref{dataMCsamples}.
The NLO+PS predictions also include photon final-state radiation and thus already part of the EW effects. Therefore no additional EW correction is applied.
 
A summary of fiducial cross-section predictions for \WW production is given in Table~\ref{tab:fidxs}.
Predictions from the different generators matched to parton showers agree well among each other and with the fixed-order predictions.
For the \sherpa\ 2.2.2 prediction, scale uncertainties are larger than for the \POWHEGBOX predictions because the \sherpa\ calculation includes matrix elements with higher jet multiplicities, which results in a larger uncertainty estimate when varying the renormalization and factorization scales in the matrix element calculation. For fixed-order predictions, scale uncertainties are large because they are evaluated according to Ref.~\cite{Stewart:2011cf}.

\begin{table}[t!]
\caption{Predictions of the \WW\ fiducial cross-section.
Predictions matched to parton showers are normalized to inclusive fixed-order calculations.}
\centering
\setlength\tabcolsep{2pt}
\begin{tabular}{@{}lllll@{}}
\toprule
Prediction & Reference & ~Normalization  &\multicolumn{2}{l}{$\sigma_\text{fiducial}$ [fb]}\\
\midrule
MATRIX NNLO \qqtoWW and \ggtoWW @ LO&\cite{Grazzini:2017mhc,Cascioli:2011va,Gehrmann:2015ora}  &~$-$ &357 &  $\pm$ 20  \\
MATRIX NNLO \qqtoWW and \ggtoWW @ NLO&\cite{Caola:2016trd}  &~$-$ & 368 &$\pm$ 21\\
(MATRIX NNLO $q\bar q$ and $gg$ @ NLO) $\times$ NLO EW&\cite{Biedermann:2016guo} & ~$-$ & 347 &$\pm$ 20  \\
\midrule
\sherpa~2.1.1~+~\openloops~\ggtoWW&\cite{jetvetoNNLO} & ~NLO~\cite{Caola:2016trd}  & \hphantom{0}19.0 &$\pm$ \hphantom{0}1.9 \\
\midrule
\POWHEGBOX+~\PYTHIAV{8}~\qqtoWW (+ \textsc{Sh.+OL} \ggtoWW)&\cite{Nason:2004rx, Frixione:2007vw, Alioli:2010xd, Melia:2011tj, Nason:2013ydw, Sjostrand:2014zea} & ~NNLO~\cite{Grazzini:2017mhc,Cascioli:2011va,Gehrmann:2015ora}~ & 350   &$\pm$ \hphantom{0}7  \\
\POWHEGBOX+~\herwigpp~\qqtoWW (+ \textsc{Sh.+OL} \ggtoWW)&\cite{Nason:2004rx, Frixione:2007vw, Alioli:2010xd, Melia:2011tj, Nason:2013ydw, Bahr:2008pv} & ~NNLO~\cite{Grazzini:2017mhc,Cascioli:2011va,Gehrmann:2015ora}& 357    &$\pm$ 11 \\
\SHERPAV{2.2.2}~\qqtoWW (+ \textsc{Sh.+OL} \ggtoWW)&\cite{Sherpa} & ~NNLO~\cite{Grazzini:2017mhc,Cascioli:2011va,Gehrmann:2015ora} & 341   &$\pm$ 20\\
\bottomrule
\end{tabular}
\label{tab:fidxs}
\end{table}
 
 
\section{Results}
\label{sec:results}
 
\subsection{Cross-section measurements and comparisons with theoretical predictions}
 
The measured fiducial cross-section for \WWemu production at $\sqrt{s}=13\,\tev$ is:
\begin{linenomath*}
$$\sigma_\text{fid}=\left(379.1\pm5.0\text{ (stat)}\pm25.4\text{ (syst)} \pm 8.0\text{ (lumi)} \right) \text{fb}~.$$
\end{linenomath*}
The combined statistical and systematic uncertainty of the measurement,
including the uncertainty in the luminosity, is 7.1\%.
 
\begin{figure}[t!]
\centering
\includegraphics[width=0.55\textwidth]{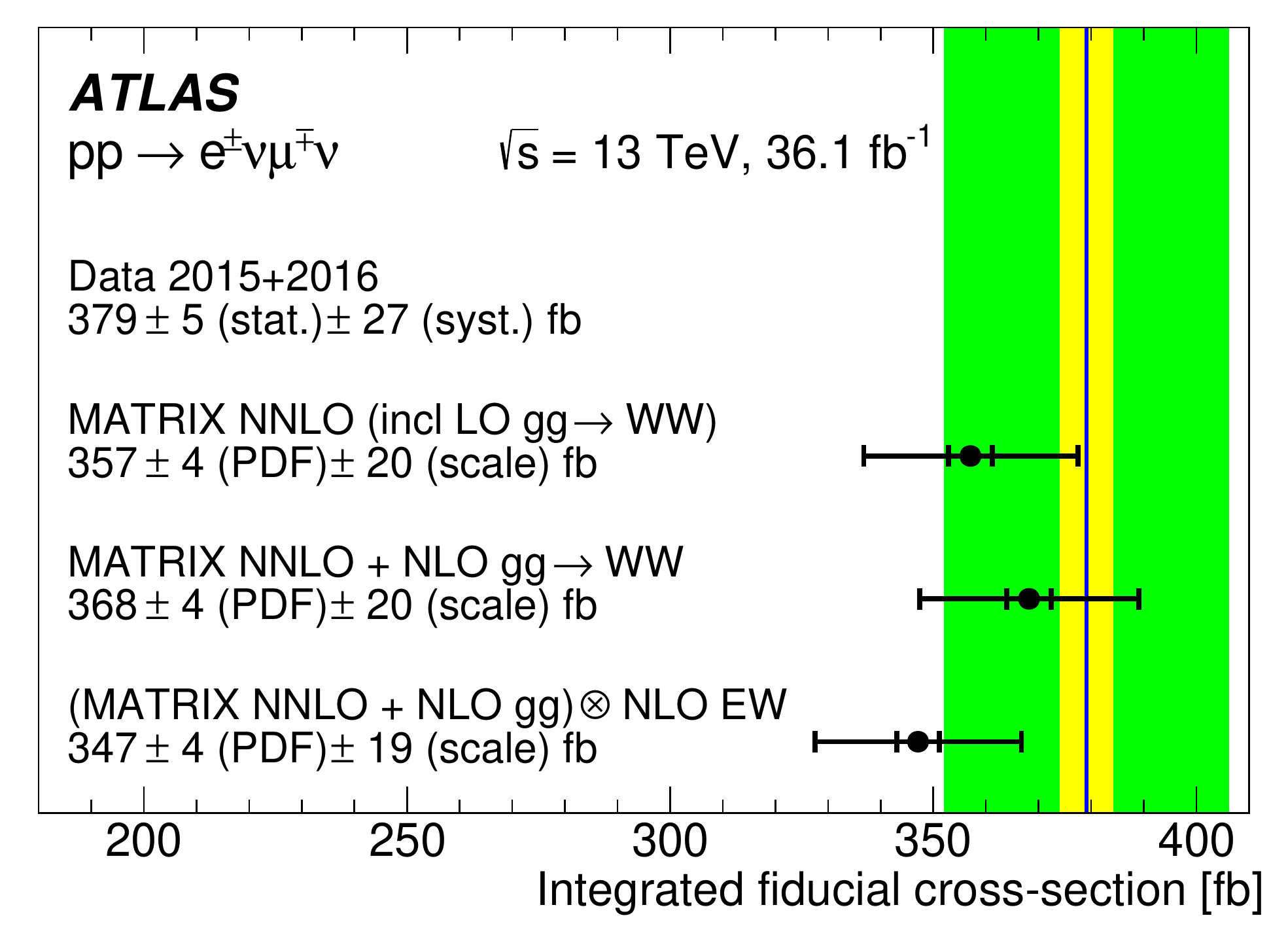}
\caption{Comparison of the measured fiducial cross-section with various theoretical predictions.
Theoretical predictions are indicated as points with inner (outer) error bars denoting PDF (PDF+scale) uncertainties.
The central value of the measured cross-section is indicated by a vertical line with the narrow band showing the statistical uncertainty and the wider band the total uncertainty including statistical and systematic uncertainties.}
\label{fig:results_fidxsec}
\end{figure}

A comparison between the fiducial cross-section measurement and fixed-order theoretical calculations is shown in \figref{results_fidxsec}.
The measurement is compared with the NNLO QCD MATRIX predictions including the full set of QCD and EW corrections, discussed in detail in \secref{theory}. The predictions agree well with the measurement.

\begin{figure}[t]
\centering
\includegraphics[width=0.49\textwidth]{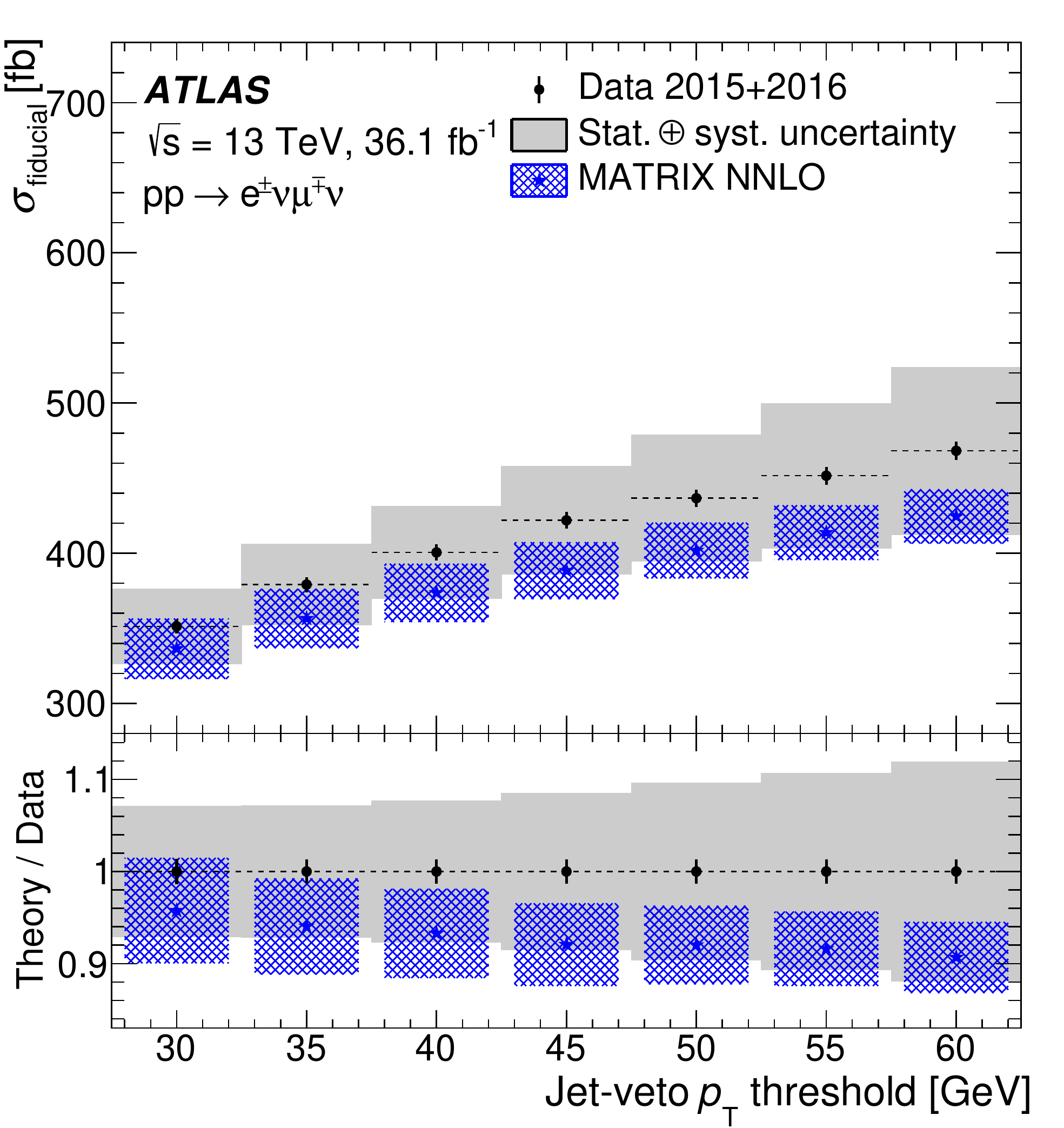}
\includegraphics[width=0.49\textwidth]{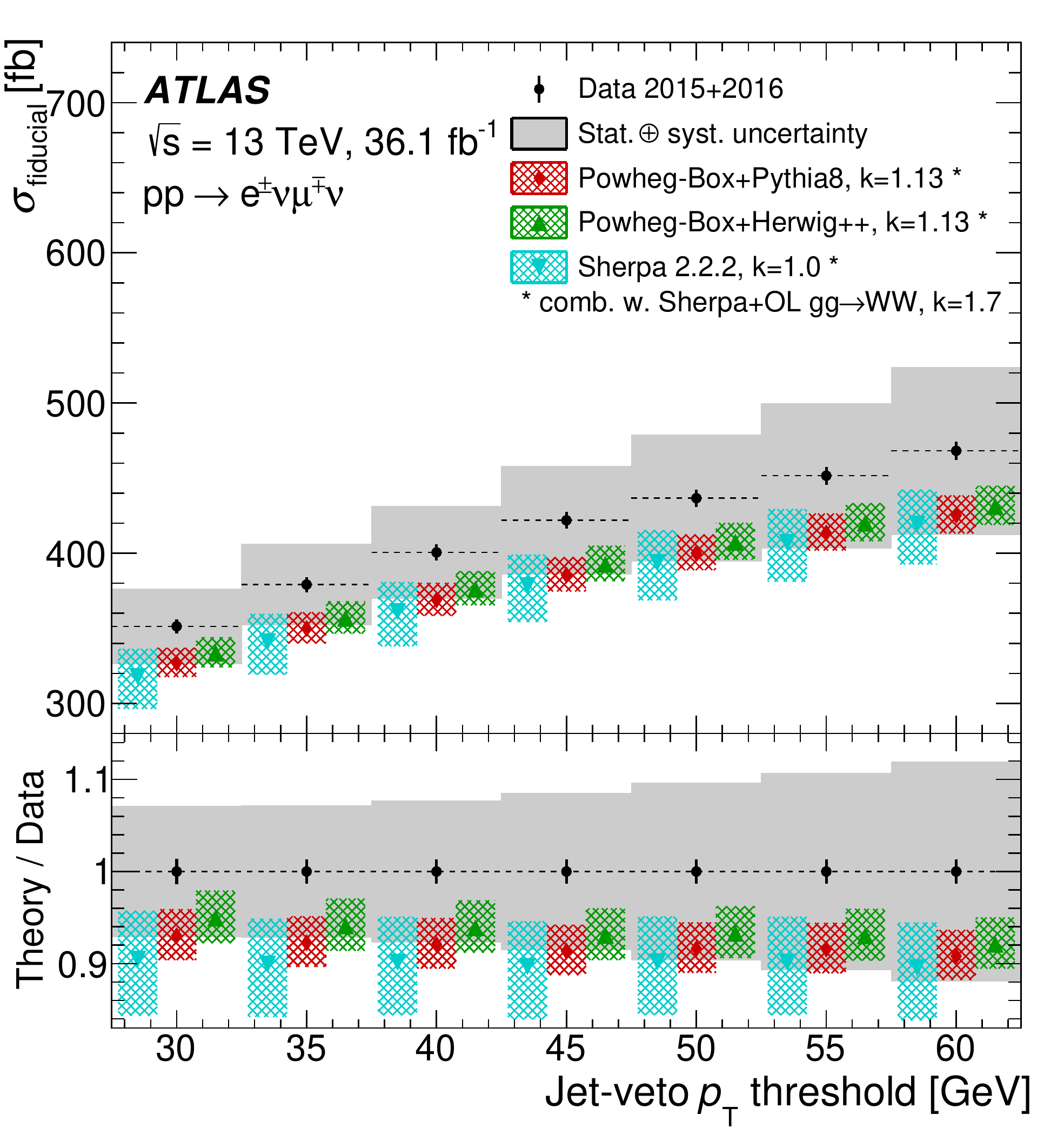}
\caption{Comparison of the measured fiducial cross-section as a function of the jet-veto \pt threshold with various theoretical predictions. The measurement is compared with NNLO predictions from MATRIX on the left. This calculation does not include the NLO EW correction and is Born-level, whilst the measurement is conducted using dressed leptons, which might account for some of the differences seen. On the right a comparison with NLO+PS predictions from \POWHEGBOX{}+\PYTHIAV{8}, \POWHEGBOX{}+\herwigpp\ and \SHERPAV{2.2.2} for \qqbar initial states, combined with \SHERPA{}+\openloops\ (LO+PS) for the $gg$ initial states is shown. All three \qqbar NLO+PS predictions are normalized to the NNLO theoretical prediction for the total cross-section, with the $gg$ LO+PS contribution normalized to NLO.
The measured cross-section values are shown as points with error bars giving the statistical uncertainty and solid bands indicating the size of the total uncertainty.
Theoretical predictions are indicated as markers with hatched bands denoting PDF+scale uncertainties.
}
\label{fig:results_fidxsecJetVeto}
\end{figure}

The measured fiducial cross-sections as a function of the jet-veto \pt thresholds are shown in \figref{results_fidxsecJetVeto}.
The fiducial cross-section rises by about 30\% when accepting events containing jets with a transverse momentum of up to 60~\gev, as compared with 30~\gev.
The measurement is compared to NNLO predictions from MATRIX (\figref{results_fidxsecJetVeto}, left), and to NLO+PS predictions from \POWHEGBOX{}+\PYTHIAV{8}, \POWHEGBOX{}+\herwigpp\ and \SHERPAV{2.2.2} for \qqbar-initiated states, combined with \SHERPA{}+\openloops\ (LO+PS) for the $gg$ initial states (\figref{results_fidxsecJetVeto}, right). All three \qqbar NLO+PS predictions are normalized to the NNLO theoretical prediction for the total cross-section, with the $gg$ LO+PS contribution normalized to NLO.
With increasing jet-veto \pt threshold, the fiducial cross-section rises as it becomes more inclusive. All predictions agree within uncertainties with the data, but are consistently at the lower bound of these.

\begin{figure}[htp]
\centering
\includegraphics[width=0.49\textwidth]{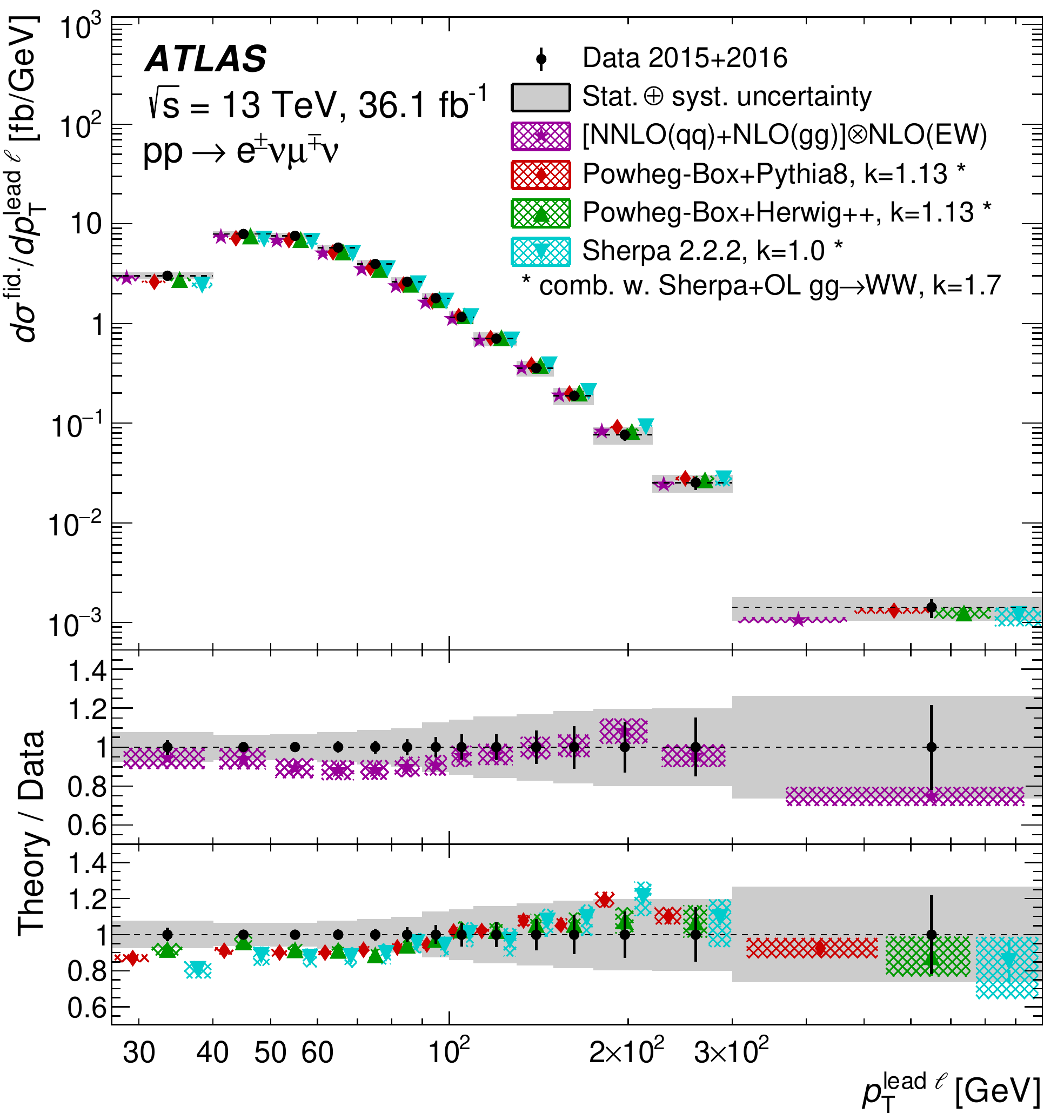}
\includegraphics[width=0.49\textwidth]{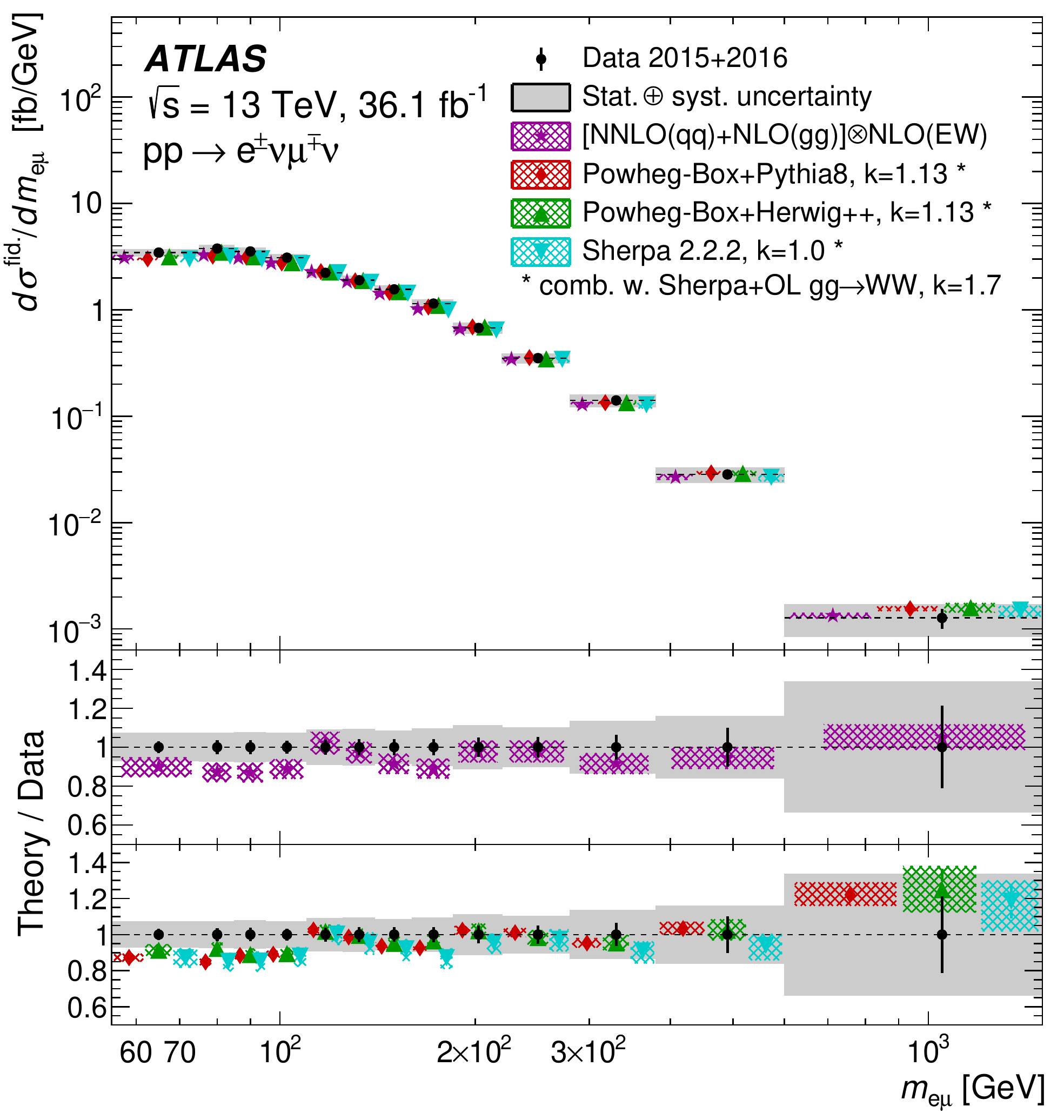}
 
\includegraphics[width=0.49\textwidth]{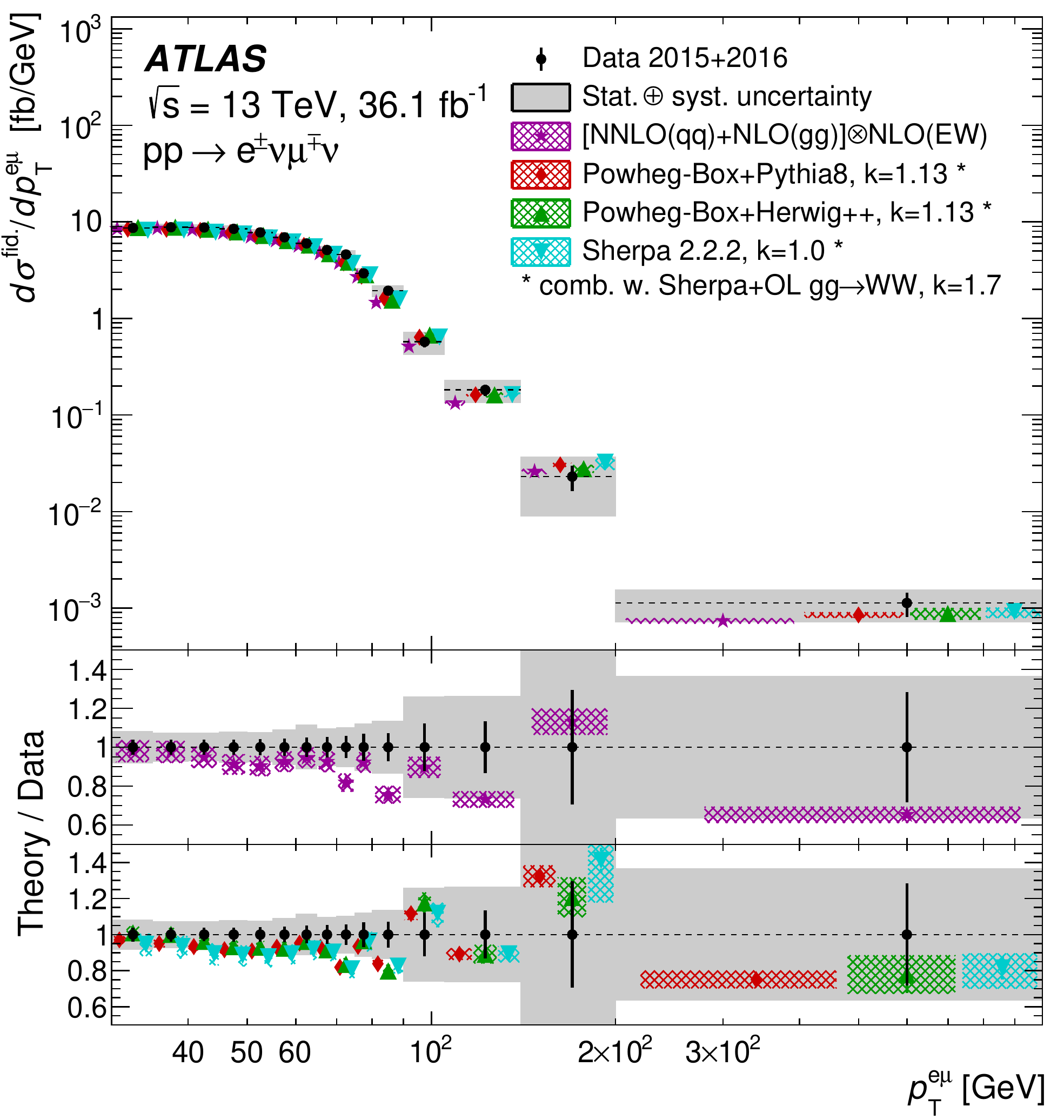}
\includegraphics[width=0.49\textwidth]{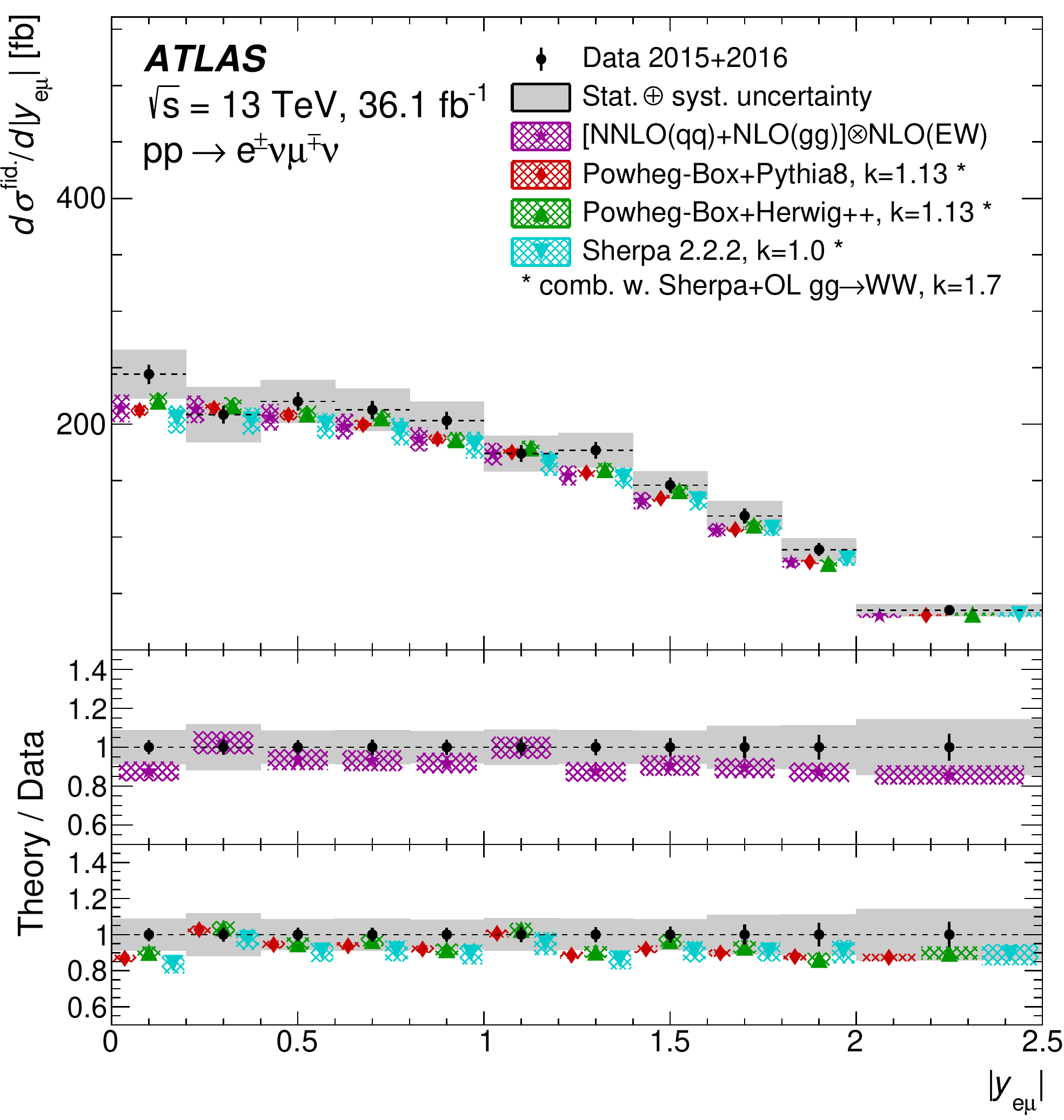}
 
\caption{Measured fiducial cross-sections of \WWemu production for four of the six observables (from left to right and top to bottom): \ptlzero, \mll, \ptll, and \yll.
The measured cross-section values are shown as points with error bars giving the statistical uncertainty and solid bands indicating the size of the total uncertainty.
The results are compared with the NNLO prediction with extra NLO EW corrections and NLO corrections for $gg\rightarrow \WW$ production, and with NLO+PS predictions from \POWHEGBOX{}+\PYTHIAV{8}, \POWHEGBOX{}+\herwigpp\ and \SHERPAV{2.2.2} for \qqbar initial states, combined with \SHERPA{}+\openloops\ (LO+PS) for the $gg$ initial states.
All three \qqbar NLO+PS predictions are normalized to the NNLO theoretical prediction for the total cross-section, with the $gg$ LO+PS contribution normalized to NLO.
Theoretical predictions are indicated as markers with hatched bands denoting PDF+scale uncertainties.}
\label{fig:results_diffxsec}
\end{figure}

\begin{figure}[htp]
\centering
\includegraphics[width=0.49\textwidth]{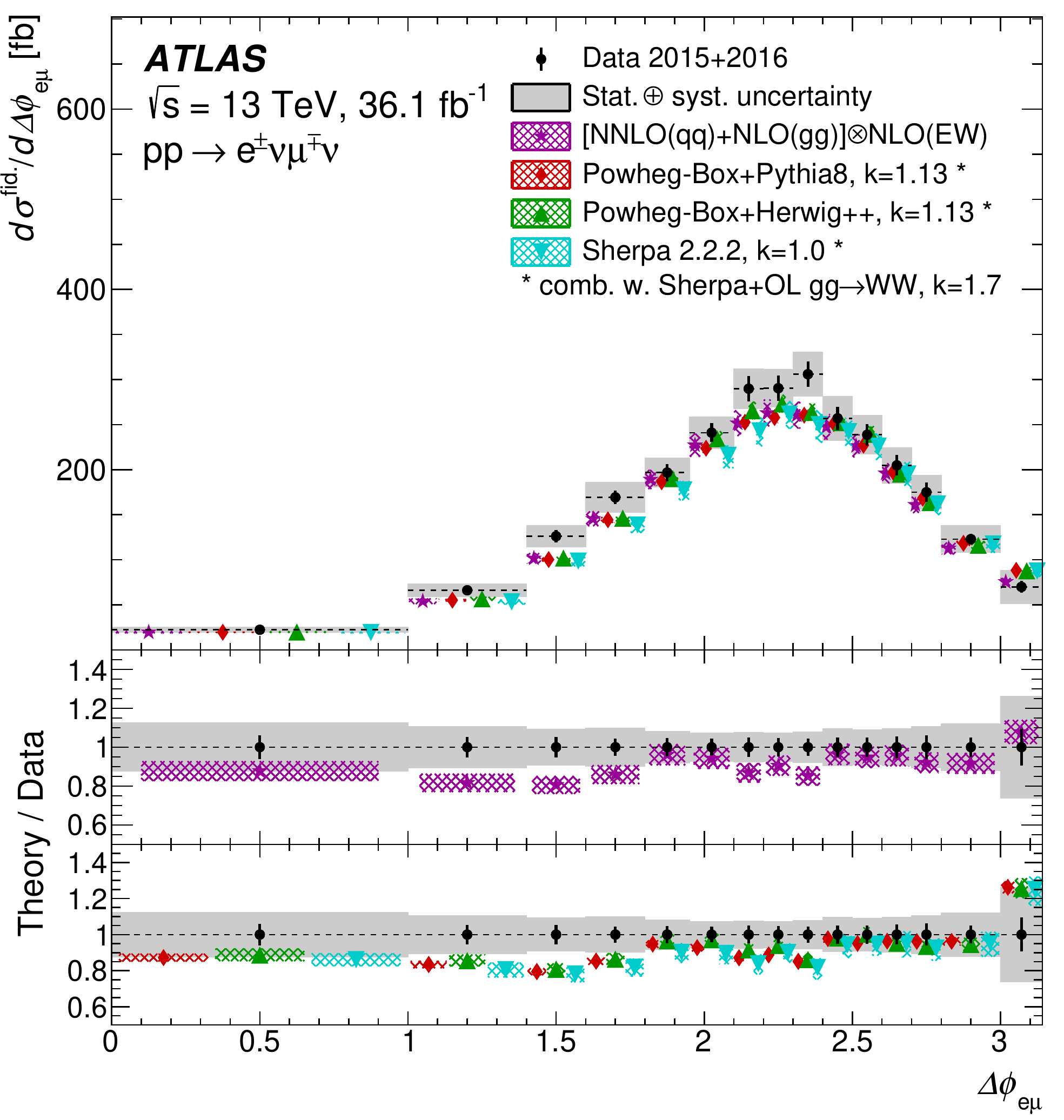}
\includegraphics[width=0.49\textwidth]{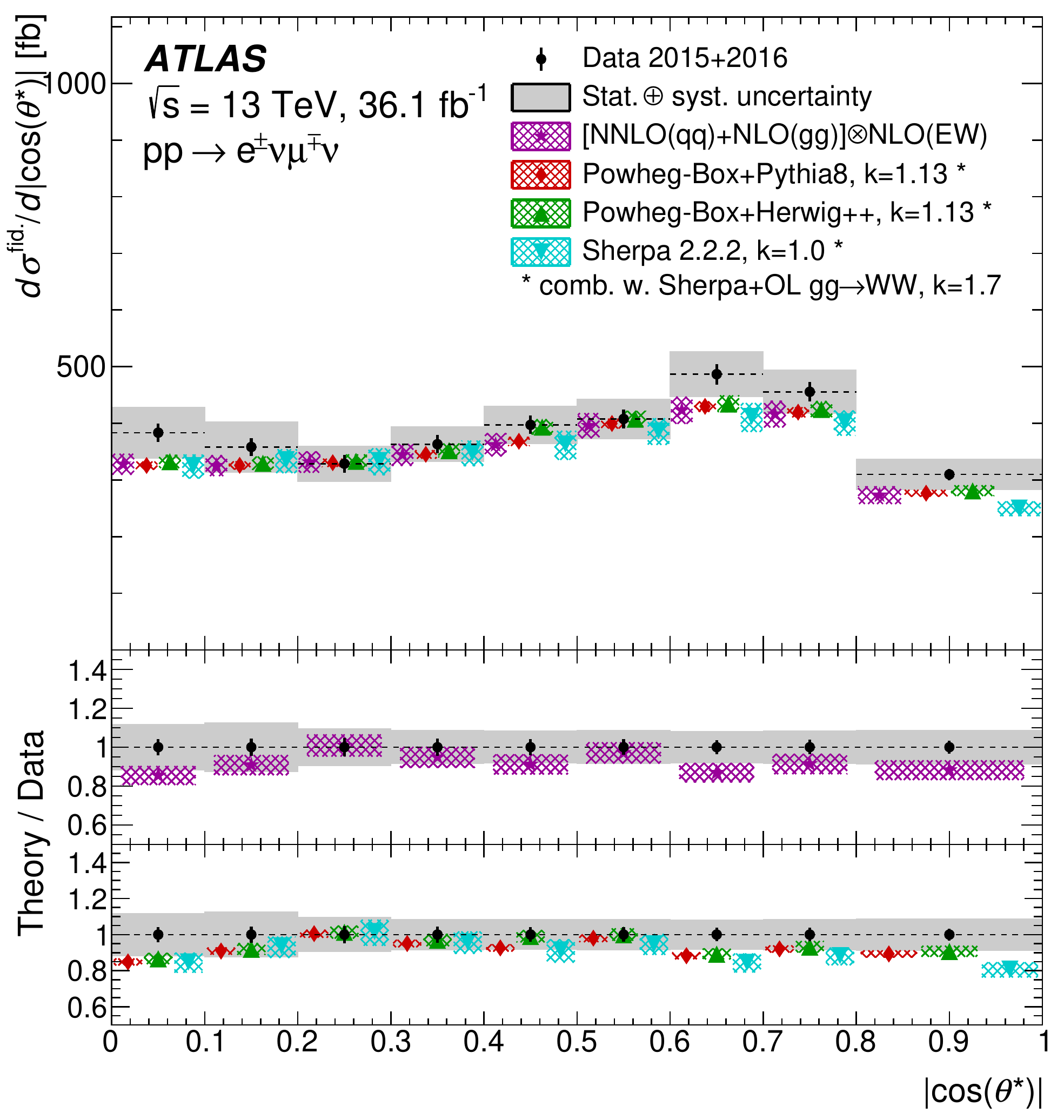}
\caption{Measured fiducial cross-sections of \WWemu production for two of the six observables: \dphill and \costheta.
The measured cross-section values are shown as points with error bars giving the statistical uncertainty and solid bands indicating the size of the total uncertainty.
The results are compared with the NNLO prediction with extra NLO EW corrections and NLO corrections for $gg\rightarrow \WW$ production, and with NLO+PS predictions from \POWHEGBOX{}+\PYTHIAV{8}, \POWHEGBOX{}+\herwigpp\ and \SHERPAV{2.2.2} for \qqbar initial states, combined with \SHERPA{}+\openloops\ (LO+PS) for the $gg$ initial states.
All three \qqbar NLO+PS predictions are normalized to the NNLO theoretical prediction for the total cross-section, with the $gg$ LO+PS contribution normalized to NLO.
Theoretical predictions are indicated as markers with hatched bands denoting PDF+scale uncertainties.}
\label{fig:results_diffxsec2}
\end{figure}

The measured fiducial cross-sections as a function of \ptlzero, \mll, \ptll, \yll, \dphill and \costheta are shown in Figures~\ref{fig:results_diffxsec} and~\ref{fig:results_diffxsec2}.
They are compared with the NNLO QCD predictions from MATRIX, including NLO corrections for $gg\rightarrow WW$ production and extra NLO EW corrections,
as well as with the same \qqbar NLO+PS predictions as stated above (combined with $gg$ LO+PS) normalized to the NNLO (NLO) theoretical prediction for the total cross-section.
All of these predictions provide a fair description of the data,
except for low values of the \pt\ of the leading lepton as well as low values of invariant mass \mll and $\dphill<1.8$.
For the \ptlzero distribution, \POWHEGBOX{}+\PYTHIAV{8} and \SHERPAV{2.2.2} underestimate the cross-section by up to 15\%--20\%.
For the other two distributions, all predictions display similar underestimates of the measured differential cross-section but to slightly varying degrees, depending on the size of their uncertainties. The most consistent difference is observed at around $\dphill\approx1.5$. A similar underprediction of the data, shifted slightly to lower \dphill values (around $\approx0.5$--$1.0$) was seen in both the ATLAS and CMS measurements at 8~\tev~\cite{STDM-2013-07,Khachatryan:2015sga} when compared with the predictions from a variety of MC generators.
Global $\chi^2$ comparisons are carried out for all the predictions. They do not display any significant differences between predictions and data with the largest $\chi^2$ per degree of freedom being 18.5/14 when comparing the \SHERPAV{2.2.2}+\SHERPA{}+\openloops\ prediction with the measured \ptlzero\ distribution.

\subsection{Limits on anomalous gauge couplings}
 
The self-couplings of the electroweak gauge bosons can be probed via the $WWZ$ and $WW\gamma$ vertices, present when the $W$ bosons are produced via $s$-channel $Z/\gamma^*$ exchange, as shown in Figure~\ref{fig:Intro_WWprodFeynmanDiagrams}.
New physics processes at a high energy scale ($\Lambda$) that alter \WW\ production can be described by operators with mass dimensions larger than four in an effective field theory (EFT) framework~\cite{EFT}. The higher-dimensional operators of the lowest order from purely EW processes have dimension six, and can generate anomalous triple-gauge-boson couplings (aTGC).
A deviation from the SM in measured \WW\ production rates or in certain kinematic distributions, as predicted by these theories, could provide evidence for physics beyond the SM.
In the EFT framework employed, there are five dimension-six operators ($\mathcal{O}_i$) and the relevant EFT coefficients (coupling constants) are: $c_{WWW}$, $c_{W}$, $c_{B}$, $c_{\tilde{W}WW}$ and $c_{\tilde{W}}$~\cite{EFT}.
The dimensionless coefficients ($c_i$) parameterize the strength of the coupling between new physics and SM particles
\begin{linenomath*}
\begin{equation*}
\mathcal{L} = \mathcal{L}_{\textrm{SM}}+ \sum_i^{}  \frac{c_i}{\Lambda^2}\mathcal{O}_i~.
\end{equation*}
\end{linenomath*}
 
Constraints on the EFT coefficients are determined by considering only one operator at a time using the unfolded leading lepton \pt\ (\ptlzero) fiducial cross-section, which was identified as the unfolded distribution most sensitive to the effect of the five operators.
 
Templates of the \ptlzero\ distribution representing the pure SM contribution, the aTGC contribution, and the interference between the SM and aTGC contributions at LO are prepared  at generator level using \MGMCatNLO version 2.6.3.2~\cite{Alwall:2011uj}, interfaced to \PYTHIAV{8.212} with the \afourteen\ tune for parton showering and hadronization. The relative size of the SM cross-section modification increases with \ptlzero\ so that the last measured bin is most sensitive to the aTGC effects. To ensure a good agreement of the \MGMCatNLO\ prediction with the baseline SM prediction, a bin-wise correction, determined as the ratio of the pure SM contributions from \POWHEGBOX{}+\PYTHIAV{8} (normalized to the NNLO cross-section) and \MGMCatNLO, is applied.
 
It is verified that the pure SM assumption used in the unfolding procedure introduces no bias to the extraction of limits from the unfolded cross-section.
A reweighting procedure implemented in the \MGMCatNLO~\cite{Frixione:2002ik} generator is used to obtain multiple signal predictions that include aTGCs of a magnitude corresponding to the upper limits set by the Run~1 analysis~\cite{STDM-2013-07}.
The simulation is interfaced to \HERWIGV{6.5}~\cite{Herwig} and passed through the ATLAS detector simulation. Neither the reconstruction efficiency and the fiducial corrections nor the bin-to-bin migrations are significantly different.
 
The measured \ptlzero\ cross-section and the \MGMCatNLO prediction, interfaced to \PYTHIAV{8}, as described above, are used to construct a likelihood function,
in which statistical and systematic measurement uncertainties are modelled by a multivariate Gaussian distribution.
Systematic uncertainties in the theory prediction are considered as nuisance parameters, each constrained with a Gaussian distribution.
Since electroweak radiative effects are already partially taken into account in the parton shower of the \MGMCatNLO prediction, the effect of applying NLO EW corrections to the \ptlzero\ distribution in addition is considered as a further systematic uncertainty.
 
Frequentist confidence intervals for the EFT coefficients are computed from values of a profile likelihood ratio test statistic~\cite{Feldman:1997qc}.
Observed and expected 95\% CL intervals for the EFT coefficients are summarized in Table~\ref{tab:eft_constraints}.
Due to the higher centre-of-mass energy, the limits reported here are more restrictive than those previously published by the ATLAS and CMS Collaborations in the \WW\ final state~\cite{STDM-2013-07, Khachatryan:2015sga}.
Compared to results from inclusive $WZ$ production~\cite{Sirunyan:2019bez} and electroweak $W$ and $Z$ boson production in association with two jets~\cite{Sirunyan:2019dyi}, both at $\sqrt{s}=13~\tev$, the limits on $c_{B}/\Lambda^2$ from this analysis are the most stringent (by about a factor 2), while those on $c_{WWW}/\Lambda^2$ and $c_{W}/\Lambda^2$ are weaker by factors of about 1.6 -- 4. Limits on the CP-odd operators $\mathcal{O}_{\tilde{W}WW}$ and $\mathcal{O}_{\tilde{W}}$ are not provided by the other two measurements.
 
The sensitivity to dimension-six operators mostly stems from their direct effect on the \WW\ cross-section as a function of \ptlzero,
except for the $c_{W}$ coefficient where both the direct contribution and the interference between the SM and terms containing EFT operators contribute equally.
 
\begin{table}[t]
\caption{
The expected and observed 95\% CL intervals for the anomalous coupling parameters of the EFT model~\cite{EFT}. There is a change in convention relative to Ref.~\cite{STDM-2013-07} that changes the sign on some of these parameters.
}
\begin{center}
\renewcommand\arraystretch{1.2}
\begin{tabular}{lcc}
\toprule
Parameter & Observed 95\% CL [\TeV$^{-2}$] & Expected 95\% CL [\TeV$^{-2}$]\\
\midrule
$c_{WWW}/\Lambda^2$ &  [ $-$3.4 , 3.3 ] & [ $-$3.0 , 3.0 ] \\
$c_{W}/\Lambda^2$ &  [ $-$7.4 , 4.1 ] &[ $-$6.4 , 5.1 ]\\
$c_{B}/\Lambda^2$ &  [ $-$21 , 18 ] &[ $-$18 , 17 ]\\
$c_{\tilde{W}WW}/\Lambda^2$ &  [ $-$1.6 , 1.6 ] & [ $-$1.5 , 1.5 ] \\
$c_{\tilde{W}}/\Lambda^2$ &  [ $-$76 , 76 ] &[ $-$91 , 91 ] \\
\bottomrule
\end{tabular}
\end{center}
\label{tab:eft_constraints}
\end{table}
 
 
\FloatBarrier
 
\section{Conclusion}
\label{sec:conclusion}
 
The cross-section for the production of \WpWm pairs in $pp$ collisions at $\sqrt{s}=13$~\tev\ (with subsequent decays into $WW\rightarrow e\nu_e\mu\nu_\mu$) is measured in a fiducial phase space that excludes the presence of jets with transverse momentum above 35~\gev.
The measurement is performed with data recorded by the ATLAS experiment at the LHC in 2015 and 2016, which correspond to an integrated luminosity of $36.1\ \text{fb}^{-1}$.
The measured fiducial cross-section is
$\sigma_\text{fid}=\left(
379.1\pm5.0\text{ (stat)}\pm25.4\text{ (syst)}\pm 8.0\text{ (lumi)}
\right) \text{fb}$,
and is found to be consistent with theoretical predictions, including NNLO QCD and NLO EW corrections.
The fiducial cross-section is also measured as a function of the transverse momentum threshold for the jet veto.
Differential cross-sections are measured as a function of kinematic and angular variables of the final-state charged leptons and are compared with several predictions from perturbative QCD calculations.
Data and theory show fair agreement for all differential distributions.
The distribution of the transverse momentum of the leading lepton is used to investigate anomalous triple-gauge-boson coupling parameters.
No evidence for anomalous $WWZ$ and $WW\gamma$ couplings is found, hence limits on their magnitudes are set.
These limits are
more restrictive than those derived at $\sqrt{s}=8~\TeV$.

\section*{Acknowledgements}
 
 
We thank CERN for the very successful operation of the LHC, as well as the
support staff from our institutions without whom ATLAS could not be
operated efficiently.
 
We acknowledge the support of ANPCyT, Argentina; YerPhI, Armenia; ARC, Australia; BMWFW and FWF, Austria; ANAS, Azerbaijan; SSTC, Belarus; CNPq and FAPESP, Brazil; NSERC, NRC and CFI, Canada; CERN; CONICYT, Chile; CAS, MOST and NSFC, China; COLCIENCIAS, Colombia; MSMT CR, MPO CR and VSC CR, Czech Republic; DNRF and DNSRC, Denmark; IN2P3-CNRS, CEA-DRF/IRFU, France; SRNSFG, Georgia; BMBF, HGF, and MPG, Germany; GSRT, Greece; RGC, Hong Kong SAR, China; ISF and Benoziyo Center, Israel; INFN, Italy; MEXT and JSPS, Japan; CNRST, Morocco; NWO, Netherlands; RCN, Norway; MNiSW and NCN, Poland; FCT, Portugal; MNE/IFA, Romania; MES of Russia and NRC KI, Russian Federation; JINR; MESTD, Serbia; MSSR, Slovakia; ARRS and MIZ\v{S}, Slovenia; DST/NRF, South Africa; MINECO, Spain; SRC and Wallenberg Foundation, Sweden; SERI, SNSF and Cantons of Bern and Geneva, Switzerland; MOST, Taiwan; TAEK, Turkey; STFC, United Kingdom; DOE and NSF, United States of America. In addition, individual groups and members have received support from BCKDF, CANARIE, CRC and Compute Canada, Canada; COST, ERC, ERDF, Horizon 2020, and Marie Sk{\l}odowska-Curie Actions, European Union; Investissements d' Avenir Labex and Idex, ANR, France; DFG and AvH Foundation, Germany; Herakleitos, Thales and Aristeia programmes co-financed by EU-ESF and the Greek NSRF, Greece; BSF-NSF and GIF, Israel; CERCA Programme Generalitat de Catalunya, Spain; The Royal Society and Leverhulme Trust, United Kingdom.
 
The crucial computing support from all WLCG partners is acknowledged gratefully, in particular from CERN, the ATLAS Tier-1 facilities at TRIUMF (Canada), NDGF (Denmark, Norway, Sweden), CC-IN2P3 (France), KIT/GridKA (Germany), INFN-CNAF (Italy), NL-T1 (Netherlands), PIC (Spain), ASGC (Taiwan), RAL (UK) and BNL (USA), the Tier-2 facilities worldwide and large non-WLCG resource providers. Major contributors of computing resources are listed in Ref.~\cite{ATL-GEN-PUB-2016-002}.
 

\clearpage

\printbibliography
 
\clearpage
 
\begin{flushleft}
{\Large The ATLAS Collaboration}

\bigskip

M.~Aaboud$^\textrm{\scriptsize 35d}$,    
G.~Aad$^\textrm{\scriptsize 101}$,    
B.~Abbott$^\textrm{\scriptsize 128}$,    
D.C.~Abbott$^\textrm{\scriptsize 102}$,    
O.~Abdinov$^\textrm{\scriptsize 13,*}$,    
A.~Abed~Abud$^\textrm{\scriptsize 70a,70b}$,    
D.K.~Abhayasinghe$^\textrm{\scriptsize 93}$,    
S.H.~Abidi$^\textrm{\scriptsize 167}$,    
O.S.~AbouZeid$^\textrm{\scriptsize 40}$,    
N.L.~Abraham$^\textrm{\scriptsize 156}$
% ...
\bigskip
\\

% ...
$^{13}$Institute of Physics, Azerbaijan Academy of Sciences, Baku; Azerbaijan.\\
% ...
$^{35}$$^{(a)}$Facult\'e des Sciences Ain Chock, R\'eseau Universitaire de Physique des Hautes Energies - Universit\'e Hassan II, Casablanca;$^{(b)}$Facult\'{e} des Sciences, Universit\'{e} Ibn-Tofail, K\'{e}nitra;$^{(c)}$Facult\'e des Sciences Semlalia, Universit\'e Cadi Ayyad, LPHEA-Marrakech;$^{(d)}$Facult\'e des Sciences, Universit\'e Mohamed Premier and LPTPM, Oujda;$^{(e)}$Facult\'e des sciences, Universit\'e Mohammed V, Rabat; Morocco.\\
% ...
$^{40}$Niels Bohr Institute, University of Copenhagen, Copenhagen; Denmark.\\
% ...
$^{70}$$^{(a)}$INFN Sezione di Pavia;$^{(b)}$Dipartimento di Fisica, Universit\`a di Pavia, Pavia; Italy.\\
% ...
$^{93}$Department of Physics, Royal Holloway University of London, Egham; United Kingdom.\\
% ...
$^{101}$CPPM, Aix-Marseille Universit\'e, CNRS/IN2P3, Marseille; France.\\
$^{102}$Department of Physics, University of Massachusetts, Amherst MA; United States of America.\\
% ...
$^{128}$Homer L. Dodge Department of Physics and Astronomy, University of Oklahoma, Norman OK; United States of America.\\
% ...
$^{156}$Department of Physics and Astronomy, University of Sussex, Brighton; United Kingdom.\\
% ...
$^{167}$Department of Physics, University of Toronto, Toronto ON; Canada.\\
% ...
$^{a}$ Also at Borough of Manhattan Community College, City University of New York, New York NY; United States of America.\\
$^{b}$ Also at Centre for High Performance Computing, CSIR Campus, Rosebank, Cape Town; South Africa.\\
$^{c}$ Also at CERN, Geneva; Switzerland.\\
$^{d}$ Also at CPPM, Aix-Marseille Universit\'e, CNRS/IN2P3, Marseille; France.\\
$^{e}$ Also at D\'epartement de Physique Nucl\'eaire et Corpusculaire, Universit\'e de Gen\`eve, Gen\`eve; Switzerland.\\
% ...
$^{*}$ Deceased

\end{flushleft}


\end{document}